\documentclass[preprintnumbers,nofootinbib,showpacs,floatfix]{revtex4}

\usepackage{graphicx}
\usepackage{latexsym}
\usepackage[english]{babel}
\usepackage{amsmath}
\usepackage{amsthm}
\usepackage{epsfig}
\usepackage{color}
\usepackage{amssymb}
\usepackage{epsfig}
\usepackage{verbatim}
\usepackage{psfrag}
\usepackage{bm}
\usepackage{bbm}
\usepackage{amssymb}

\begin{document}

\title{\Large Decoherence in Quantum Mechanics}

\preprint{ITP-UU-10/45, SPIN-10/38}

\preprint{HD-THEP-10-23 }


\author{Jurjen F. Koksma}
\affiliation{Institute for Theoretical Physics (ITP) \& Spinoza
Institute, Utrecht University, Postbus 80195, 3508 TD Utrecht, The
Netherlands \\ \texttt{\textup{J.F.Koksma@uu.nl}}}

\author{Tomislav Prokopec}
\affiliation{Institute for Theoretical Physics (ITP) \& Spinoza
Institute, Utrecht University, Postbus 80195, 3508 TD Utrecht, The
Netherlands \\ \texttt{\textup{T.Prokopec@uu.nl}}}

\author{Michael G. Schmidt}
\affiliation{Institut f\"ur Theoretische Physik, Heidelberg
University, Philosophenweg 16, D-69120 Heidelberg, Germany
\\ \texttt{\textup{M.G.Schmidt@thphys.uni-heidelberg.de}}}

\begin{abstract}
We study decoherence in a simple quantum mechanical model using
two approaches. Firstly, we follow the conventional approach to
decoherence where one is interested in solving the reduced density
matrix from the perturbative master equation. Secondly, we
consider our novel correlator approach to decoherence where
entropy is generated by neglecting observationally inaccessible
correlators. We show that both methods can accurately predict
decoherence time scales. However, the perturbative master equation
generically suffers from instabilities which prevents us to
reliably calculate the system's total entropy increase. We also
discuss the relevance of the results in our quantum mechanical
model for interacting field theories.
\end{abstract}

\maketitle

\section{Introduction}
\label{Introduction}

\subsection{The Correlator Approach to Decoherence}
\label{The Correlator Approach to DecoherenceINTRO}

Recently, we proposed a novel approach to decoherence based on
neglecting observationally inaccessible correlators
\cite{Koksma:2009wa, Koksma:2010zi}, also see
\cite{Giraud:2009tn}. Just as in the conventional approach to
decoherence \cite{Zeh:1970, Zurek:1981xq, Joos:1984uk,
Zurek:1991vd}, we split our world into a distinct system,
environment, and observer. Interactions between system and
environment in quantum field theory generate non-Gaussian
correlators. However, from the observer's perspective, these
correlators are generically small in the perturbative sense. The
observer's inability to measure the information stored in these
higher order correlators generates a certain amount of entropy
which turns our quantum system into a classical stochastic system.
We applied our ``correlator approach'' to decoherence to an
interacting quantum field theoretical model \cite{Koksma:2009wa},
with an interaction term of the form $h \phi(x)\chi^2(x)$, by
solving for the renormalised statistical propagator in an
out-of-equilibrium setting. Giraud and Serreau
\cite{Giraud:2009tn} performed a similar analysis in a $\lambda
\phi^4(x)$ field theory. Starting point is the well known von
Neumann entropy:
\begin{equation} \label{vNeumannEntropy}
S_{\rm vN} = - {\rm Tr}[\hat \rho\ln(\hat \rho)]\,,
\end{equation}
where $\hat\rho$ denotes the density operator which in the
Schr\"odinger picture satisfies the von Neumann equation:
\begin{equation} \label{vNeumannEquation}
\imath \hbar\frac{\partial}{\partial t}\hat \rho  = [\hat H,\hat
\rho] \,.
\end{equation}
By restricting ourselves to a Gaussian density matrix (and thus
neglecting all non-Gaussian information that is contained in the
full density matrix), one can easily relate the Gaussian von
Neumann entropy to the statistical propagator of our field
theoretical system \cite{Koksma:2010zi,Campo:2008ij}. This
procedure can be improved by including non-Gaussian corrections to
the entropy \cite{Koksma:2010zi}. Indeed, the Gaussian von Neumann
entropy is in most cases no longer conserved, unlike the total von
Neumann entropy in equation (\ref{vNeumannEntropy}).

We thus argue to solve directly for the Gaussian correlators
characterising the Gaussian properties of our system. A
sophisticated field theoretical framework exists to renormalise
the statistical propagator, include perturbative corrections and
study genuine time evolution in an out-of-equilibrium setting (see
e.g. \cite{Berges:2004yj, Garny:2009ni}).

\subsection{The Master Equation Approach to Decoherence and its Shortcomings}
\label{The Master Equation Approach to Decoherence and its
Shortcomings}

In the conventional approach to decoherence one is interested in
solving for the reduced density matrix which is obtained by
tracing over the environmental degrees of freedom in the full
density matrix:
\begin{equation} \label{reduceddensitymatrix}
\hat{\rho}_{\mathrm{red}}(t)= \mathrm{Tr}_E [\hat\rho(t)]\,.
\end{equation}
The unitary von Neumann equation (\ref{vNeumannEquation})
transforms to a non-unitary master equation for
$\hat{\rho}_{\mathrm{red}}$. The master equation is in principle
equivalent to the von Neumann equation if no approximations are
made. This is for example clear from the influence functional
method. In other words: the exact master equation is just as hard
to solve as the von Neumann equation in interacting quantum field
theories. Therefore, one usually relies on perturbative methods to
simplify the exact master equation and to derive a perturbative
master equation.

The conventional approach to decoherence suffers from both
theoretical and practical shortcomings. From a theoretical point
of view, it is disturbing that $\rho_{\mathrm{red}}$ is obtained
from a non-unitary equation despite of the fact that the
underlying theory is unitary\footnote{The energy is not conserved
as a consequence. This means that this basic method to check a
particular numerical evolution for $\rho_{\mathrm{red}}$ is not
available.}. Practically speaking, the (exact or perturbative)
master equation is still so hard to solve that even the most basic
field theoretical questions have never been properly addressed: no
well established framework exists to include perturbative
corrections to a reduced density matrix (see however the
pioneering work in quantum mechanical cases \cite{Hu:1993vs,
Hu:1993qa, Calzetta:book}) nor has any reduced density matrix ever
been renormalised\footnote{For example in \cite{Burgess:2006jn}
the decoherence of inflationary primordial fluctuations is
investigated using the master equation approach however
renormalisation is not addressed.}.

\subsection{A Clean Comparison Between the Two Approaches}
\label{A Clean Comparison Between the Two Approaches}

As a solution to the perturbative master equation in a field
theoretical setting has so far not been derived, let us study a
simple quantum mechanical model. Quantum mechanics thus provides
us with the ideal playing field to compare the two approaches to
decoherence. In this paper we study a well known textbook example
\cite{Grabert:1988yt, Paz:2000le, Zurek:2003zz, Calzetta:book} of
$N+1$ simple harmonic oscillators. Here, one oscillator, $x$,
plays the role of the system and the other $N$ oscillators, $q_n$,
$1 \leq n \leq N$, play the role of the environment. The
oscillators will be coupled quadratically: $\lambda_n x q_n$. In
quantum mechanics we do not have to worry about divergences and
moreover, since we have chosen our model to be Gaussian, we can
actually compute the reduced density matrix by tracing over the
environmental degrees of freedom. Indeed, all the practical
shortcomings to the conventional approach to decoherence as listed
above have thus been circumvented.

Furthermore, it is important to realise that the entropy in our
correlator approach to decoherence is not generated by neglecting
non-Gaussianities as our quantum mechanical model is Gaussian.
Instead, the correlation entropy between the system and
environment that builds up due to the coupling between the two is
neglected. The correlation entropy in this paper thus plays the
role of the non-Gaussian contributions to the entropy in a proper
field theory. In this paper we present two calculations:
\begin{itemize}
\item We calculate the evolution of the entropy by solving for the
reduced density matrix from the perturbative master equation;
\item We calculate the evolution of the entropy by solving for the
Gaussian correlators directly from which we calculate the Gaussian
von Neumann entropy.
\end{itemize}
In particular we identify regions in parameter space where the
perturbative master equation performs well and where it does not.
In section \ref{The Model} we introduce the model and in section
\ref{The Master Equation} we outline two derivations of the
perturbative master equation. In sections \ref{Two Coupled
Oscillators} and \ref{N Coupled Oscillators} we analyse the
evolution of the entropy for $N=1$ and $N=50$, respectively.

\section{The Model}
\label{The Model}

\subsection{Bilinearly Coupled Simple Harmonic Oscillators}
\label{Bilinearly Coupled Simple Harmonic Oscillators}

We consider the following total quantum mechanical action:
\begin{equation}\label{action:QM}
S\left[\tilde x,\{\tilde q_n\}\right] = \int \mathrm{d} t L \left
[\tilde x,\{\tilde q_n\}\right] = \int \mathrm{d} t \left\{
L_{\mathrm{S}}\left[\tilde x\right] + L_{\mathrm{E}}
\left[\{\tilde q_n\}\right] + L_{\mathrm{Int}}\left[\tilde
x,\{\tilde q_n\}\right] \right\}\,,
\end{equation}
where:
\begin{subequations}
\label{action:QM2}
\begin{eqnarray}
L_{\mathrm{S}} \left[\tilde x\right] &=& \frac12 m \dot {\tilde
x}^2 - \frac12 m \omega^2_0 \tilde x^2 \label{action:QM2a}\\
L_{\mathrm{E}} \left[\{\tilde q_n\}\right] &=&
\sum_{n=1}^N\left(\frac12
m_n \dot{\tilde q}_n^2 - \frac12 m_n \omega_n^2 \tilde q_n^2\right) \label{action:QM2b}\\
 L_{\mathrm{Int}}\left[\tilde x,\{\tilde q_n\}\right] &=& - \sum_{n=1}^N \tilde \lambda_n \tilde q_n \tilde
 x\,.
\end{eqnarray}
\end{subequations}
Here $\tilde x=\tilde x(t)$ denotes the system coordinate and
$\tilde q_n=\tilde q_n(t)$ with $n=1,\cdots,N$ are the environment
coordinates. It is useful to perform the rescaling:
\begin{subequations}
\label{rescaling}
\begin{eqnarray}
x &=& \sqrt{m}\tilde x \label{rescalinga}\\
q_n &=& \sqrt{m_n}\tilde q_n  \label{rescalingb} \,,
\end{eqnarray}
\end{subequations}
after which the Lagrangian in equation (\ref{action:QM}) becomes:
\begin{equation}\label{action:QM3}
L\big[x,\{q_n\}\big] = \frac12 {\dot x}^2 - \frac12 \omega^2_0
 x^2 + \sum_{n=1}^N\left(\frac12 {\dot q}_n^2 - \frac12
\omega_n^2 q_n^2 \right) - \sum_{n=1}^N \lambda_n q_n  x \,,
\end{equation}
where $\lambda_n = \tilde \lambda_n/\sqrt{mm_n}$. Throughout the
paper we will set $\langle \hat{x}\rangle = 0$ and $\langle
\hat{q}_n \rangle = 0$, $1 \leq n \leq N$. In order to understand
how to impose initial conditions later on, it is useful to
explicitly derive the statistical propagator in thermal
equilibrium. In thermal equilibrium the Schwinger-Keldysh
propagators for e.g. the coordinate $x$ are given by:
\begin{subequations}
\label{propagators:QMt}
\begin{eqnarray}
 \imath\Delta^{++}(t;t')&=& \frac{\cos(\omega_0\Delta t)}{2\omega_0}
                                 \coth\Big(\frac{\beta\omega_0}{2}\Big)
                         - \frac{\imath}{2}{\rm sign}(\Delta t)
                                \frac{\sin(\omega_0\Delta t)}{\omega_0}
\label{propagators:QMta}\\
 \imath\Delta^{--}(t;t')&=& \frac{\cos(\omega_0\Delta t)}{2\omega_0}
                                 \coth\Big(\frac{\beta\omega_0}{2}\Big)
                         + \frac{\imath}{2}{\rm sign}(\Delta t)
                                \frac{\sin(\omega_0\Delta t)}{\omega_0}
      =\Big[\imath\Delta^{--}(t;t')\Big]^*
\label{propagators:QMtb}\\
 \imath\Delta^{+-}(t;t')&=&  \frac{\cos(\omega_0\Delta t)}{2\omega_0}
                                 \coth\Big(\frac{\beta\omega_0}{2}\Big)
                         + \frac{\imath}{2}\frac{\sin(\omega_0\Delta t)}{\omega_0}
\label{propagators:QMtc}\\
 \imath\Delta^{-+}(t;t')&=&  \frac{\cos(\omega_0\Delta t)}{2\omega_0}
                                 \coth\Big(\frac{\beta\omega_0}{2}\Big)
                         - \frac{\imath}{2}\frac{\sin(\omega_0\Delta t)}{\omega_0}
      =\Big[\imath\Delta^{+-}(t;t')\Big]^* \label{propagators:QMtd} \,,
\end{eqnarray}
\end{subequations}
where $\beta =1/(k_BT)$, with $k_B$ the Stefan-Boltzmann constant
and where we defined $\Delta t = t-t'$. The causal and statistical
propagators now follow as the sum and the difference of the two
Wightman propagators:
\begin{subequations}
\label{propagators2}
\begin{eqnarray}
\Delta^{c}(t;t^\prime) &=& \Delta^{-+}(t;t^\prime)
                       - \Delta^{+-}(t;t^\prime)
     = -\frac{\sin(\omega_0\Delta t)}{\omega_0}
\label{propagatorcausal} \\
F(t;t^\prime) &=& \frac{1}{2}\Big(\imath\Delta^{-+}(t;t^\prime)
  +\imath\Delta^{+-}(t;t^\prime)\Big)
    =  \frac{\cos(\omega_0\Delta t)}{2\omega_0}
                                 \coth\Big(\frac{\beta\omega_0}{2}\Big)
\label{propagatorsstatistical}
\,.
\end{eqnarray}
\end{subequations}
%

\subsection{The Von Neumann Entropy in Quantum Mechanics}
\label{The Von Neumann Entropy in Quantum Mechanics}

\subsubsection{$\lambda = 0$}
\label{lambda zero}

It is important to realise that since the Lagrangian in equation
(\ref{action:QM3}) is Gaussian, no non-Gaussianities are generated
by the coupling $\lambda$. For a free quantum mechanical system
($\lambda = 0$), it is simple to find the total von Neumann
entropy in equation (\ref{vNeumannEntropy}):
\begin{equation}
S_{\mathrm{vN}} = - \mathrm{Tr} [\hat{\rho}_{\mathrm{f}} \log
\hat{\rho}_{\mathrm{f}}] \nonumber \,,
\end{equation}
where the subscript ``f'' for ``free'' reminds us that all
interactions or couplings are switched off. The density operator
of a quantum mechanical Gaussian state centered at the origin is
of the form:
\begin{subequations}
\label{density operator: particle}
\begin{equation} \label{density operator: particle1}
\hat{\rho}_{\mathrm{f}}(t) = \int_{-\infty}^{\infty} \mathrm{d}x
\int_{-\infty}^{\infty} \mathrm{d}y |x\rangle
\rho_{\mathrm{f}}(x,y;t)\langle y |\,,
\end{equation}
where:
\begin{equation} \label{density operator: particle2}
\rho_{\mathrm{f}}(x,y;t) = {\cal N}(t) \exp\left[
-a(t)x^2-b(t)y^2+2c(t)xy \right] \,,
\end{equation}
\end{subequations}
and where $a=a(t)$, $b=b(t)$ and $c=c(t)$ are determined from the
von Neumann equation. Moreover, from $\hat
\rho_{\mathrm{f}}^\dagger = \hat\rho_{\mathrm{f}}$ it follows that
$b^* = a$ and $c^*=c$. When $c\neq 0$ the density matrix is mixed
and entangled. The normalisation ${\cal N}$ is obtained from
requiring $\mathrm{Tr}[\hat{\rho}_{\mathrm{f}}]=1$:
\begin{equation}\label{Norm of rho:2}
{\cal N} =  \sqrt{ \frac{2(a_{\mathrm{R}}-c)}{\pi}} \,,
\end{equation}
provided that $c < a_{\mathrm{R}}$ where $a_{\mathrm{R}}=\Re[a]$.
We can furthermore derive the three non-trivial Gaussian
correlators characterising our system:
\begin{subequations}
\label{correlators:all}
\begin{eqnarray}
\langle \hat x^2\rangle &=& {\rm Tr}[\hat\rho_{\mathrm{f}}\hat
x^2] = \int_{-\infty}^{\infty} \mathrm{d}\tilde{x} \langle
\tilde{x}| \hat\rho_{\mathrm{f}}\hat x^2 | \tilde{x}\rangle =
\frac{1}{4(a_{\mathrm{R}} -c)}
\label{correlators:alla}\\
\Big\langle \frac12\{\hat x,\hat p\}\Big\rangle &=&
-\frac{a_{\mathrm{I}} }{2(a_{\mathrm{R}} -c)} \label{correlators:allb}\\
\langle \hat p^2\rangle &=& \frac{|a|^2-c^2}{a_{\mathrm{R}} -c}
\,. \label{correlators:allc}
\end{eqnarray}
\end{subequations}
We used $\langle x| \hat p |\psi\rangle = - \imath
\partial_{x} \langle x|\psi\rangle$. These relations can be
inverted:
\begin{subequations}
\label{aI:aR:c}
\begin{eqnarray}
\label{aI:aR:ca} a_{\mathrm{I}}  &=& -\frac{\left\langle
\frac12\{\hat x,\hat p\}\right\rangle}
                            {2\langle \hat x^2\rangle}
\label{aI:aR:cb}\\
a_{\mathrm{R}}  &=& \frac{\Delta^2+1}{8\langle \hat x^2\rangle}
\label{aI:aR:cc}\\
c &=& \frac{\Delta^2-1}{8\langle \hat x^2\rangle} \,,
\label{aI:aR:cd}
\end{eqnarray}
\end{subequations}
We can straightforwardly find the von Neumann entropy by making
use of the replica trick (see e.g. \cite{Koksma:2010zi}):
\begin{equation} \label{vNeumannEntropy2}
S_{\rm vN} = \frac{\Delta+1}{2}\ln\left(\frac{\Delta+1}{2}\right)
- \frac{\Delta-1}{2}\ln\left(\frac{\Delta-1}{2}\right)\,,
\end{equation}
where:
\begin{equation}\label{PhaseSpaceArea1}
\Delta^2  = \frac{a_{\mathrm{R}} +c}{a_{\mathrm{R}} -c}  = 4
\left[\langle \hat x^2\rangle \langle \hat p^2\rangle -
\Big\langle \frac12\{\hat x,\hat p\}\Big\rangle^2 \right] = 4
\left. \left[ F(t;t^\prime)
\partial_t
\partial_{t'} F(t;t^\prime) -
\{ \partial_t F(t;t^\prime)\}^2 \right]\right|_{t=t'} \,.
\end{equation}
The physical meaning of $\Delta/2$ is the phase space area
occupied by a Gaussian state centered at the origin in units of
$\hbar$. For a pure state we have $\Delta=1$, whereas for a mixed
state $\Delta >1$. The phase space area is an extremely important
function, as it coincides precisely with the area the state
occupies in Wigner space \cite{Koksma:2010zi} and is moreover
conserved by the evolution $\mathrm{d}/\mathrm{d}t[\Delta^2] = 0$
in free theories. It is also important to mention that we can
associate a statistical particle number to $\Delta$ given by
$n=(\Delta-1)/2$ such that the von Neumann entropy reduces to the
well-known entropy of $n$ free Bose particles per quantum state.

\subsubsection{$\lambda \neq 0$}
\label{lambda not zero}

As in all unitary systems, the von Neumann entropy following from
equation (\ref{vNeumannEntropy}) is conserved:
\begin{equation}\label{vNeumannEntropy4}
S_{\mathrm{vN}} = \mathrm{const} \,.
\end{equation}
The von Neumann entropy for the model (\ref{action:QM3}) contains
both information about the system, the environment and the
correlations between system and environment that are generated due
to the coupling:
\begin{equation}\label{vNeumannEntropy5}
S_{\mathrm{vN}} = S_{\mathrm{total}} = S_{S}(t)+
S_{E}(t)+S_{SE}(t)\,.
\end{equation}
where $S_S$ and $S_E$ represent the Gaussian von Neumann entropy
contained in the system and environment separately, and where
$S_{SE}$ denotes the Gaussian correlation entropy. Adding the word
``Gaussian'' to these entropies might at the moment seem
redundant, because our total system and environment are Gaussian
so no non-Gaussianities can be generated as we mentioned before.
We nevertheless insist on this nomenclature in order to easily
compare it with interacting field theories later. For a free
system $S$ that is not coupled to or in interaction with an
environment $E$ we have derived the von Neumann entropy in
equation (\ref{vNeumannEntropy2}) and (\ref{PhaseSpaceArea1}).
Likewise, we can associate a Gaussian von Neumann entropy to our
system $S$ in the presence of a coupling $\lambda$ too:
\begin{equation}\label{vNeumannEntropy6}
S_{S}(t) =
\frac{\Delta(t)+1}{2}\ln\left(\frac{\Delta(t)+1}{2}\right) -
\frac{\Delta(t)-1}{2}\ln\left(\frac{\Delta(t)-1}{2}\right) \,,
\end{equation}
where the phase space area is given in equation
(\ref{PhaseSpaceArea1}) and follows from measuring the three
non-trivial equal time Gaussian correlators characterising the
properties of the system: $\langle \hat{x}^2 \rangle$, $\langle
\hat{p}_x^2 \rangle$ and $1/2\langle\{\hat x,\hat p_x\}\rangle$.
Note that for the coupled system $S_{S}(t)$ is generally not
conserved. A similar argument applies to the environmental
oscillators and we can likewise associate an environmental entropy
to all $q_n$ if we measure its properties. Since the total von
Neumann entropy is conserved in equation (\ref{vNeumannEntropy5}),
we can then find the time evolution of the correlation entropy by
solving:
\begin{equation}\label{vNeumannEntropy7}
S_{SE}(t) = S_{\mathrm{vN}} - S_{S}(t) -  S_{E}(t) = S_{S}(0) +
S_{E}(0) - S_{S}(t) -  S_{E}(t) \,,
\end{equation}
where we assumed that $S_{SE}(0)=0$ initially. What we would refer
to as ``Gaussian von Neumann entropy'' is referred to as
``correlation entropy'' in \cite{Calzetta:2003dk}, where they
prove an $H$-theorem for a quantum mechanical $O(N)$ model.

The reduced density matrix captures the average effect from the
environmental oscillators on our system oscillator and is defined
in equation (\ref{reduceddensitymatrix}):
\begin{equation}
\hat{\rho}_{\mathrm{red}}(t)= \mathrm{Tr}_E
[\hat\rho(t)]\nonumber\,.
\end{equation}
In order to quantify the amount of decoherence that has taken
place, one is interested in the ``reduced von Neumann entropy''
which we can define by\footnote{This entropy is not to be confused
with the entanglement entropy of for example black holes which is
generated by the observer's inability to access the information in
all of the space-time.}:
\begin{equation}\label{reducedvNeumannEntropy}
S_{\mathrm{vN}}^{\mathrm{red}}(t) = - \mathrm{Tr}
[\hat{\rho}_{\mathrm{red}}(t) \log \hat{\rho}_{\mathrm{red}}(t)]
\,.
\end{equation}
The Gaussian von Neumann entropy we defined in equation
(\ref{vNeumannEntropy6}) is identical to the ``reduced von Neumann
entropy'' defined above:
\begin{equation}\label{entropyequality}
S_{S}(t) = S_{\mathrm{vN}}^{\mathrm{red}}(t)\,.
\end{equation}
The argument is trivial:
\begin{equation}
\langle \hat x^2\rangle = {\rm Tr}[\hat\rho\hat x^2] =
\int_{-\infty}^{\infty} \mathrm{d}\tilde{x} \mathrm{d}\tilde{q}
\langle \tilde{x}\tilde{q}| \hat\rho \hat x^2 | \tilde{q}
\tilde{x}\rangle = {\rm Tr}_S[\hat\rho_{\mathrm{red}} \hat x^2]
\label{entropyequality2}\,,
\end{equation}
and similar expressions hold for
$\langle\{\hat{x},\hat{p}_x\}\rangle$ and
$\langle\hat{p}_x^2\rangle$. Since the value of all three
non-trivial Gaussian correlators does not depend on whether we
evaluate the expectation value with the full density matrix or
with the reduced density matrix, the value of the phase space area
$\Delta(t)$ or indeed the entropy is identical in both cases. We
show this explicitly using our coupled system of simple harmonic
oscillators in appendix \ref{Reducing the Density Matrix}. Of
course, when proper (non-Gaussian) interactions are considered,
equation (\ref{entropyequality}) is only true for the ``Gaussian
reduced von Neumann entropy''
$S_{\mathrm{vN}}^{\mathrm{G,red}}(t)$ based on including Gaussian
correlators only.

\section{The Master Equation}
\label{The Master Equation}

In this section we outline for completeness two derivations of the
master equation. Paz and Zurek nicely derive the master equation
by means of perturbative methods \cite{Paz:2000le} before they
review the ``influence functional method''. The latter was
pioneered by Caldeira and Leggett \cite{Caldeira:1982iu}, see also
the early review of Grabert, Schramm and Ingold
\cite{Grabert:1988yt}, which in principle allows us to derive a
non-perturbative master equation \cite{Su:1987pi, Hu:1991di,
Hu:1993vs, Hu:1993qa, Chou:2007jg}. In the perturbative
approximation both derivations yield the same master equation.

One defines the reduced density matrix by tracing over the
environmental degrees of freedom in the full density matrix as in
equation (\ref{reduceddensitymatrix}). In most of the cases of
physical interest, it is impossible to solve for the von Neumann
equation of the density matrix exactly. Hence, it is impossible to
perform the trace to find the reduced density matrix exactly.
Consequently, in order to write down an equation of motion for
$\rho_{\mathrm{red}}$ that is usually referred to as a ``master
equation'', one makes certain approximations to the von Neumann
equation. For example, one can make a perturbative approximation
or neglect the backreaction from the system on the environment.
The master equation is non-unitary in nature which is precisely
what generates the entropy.

Let us state again explicitly at this point that there is nothing
wrong with tracing over environmental degrees of freedom per se,
it is the perturbative master equation that is used to obtain the
``reduced density matrix'' that fails to capture the correct
physics at late times as we will see later.

Let us begin by following the perturbative approach as outlined by
Paz and Zurek in e.g. \cite{Paz:2000le}. The Hamiltonian following
from (\ref{action:QM3}) is given by:
\begin{equation}\label{HamiltonianQM1}
\hat{H} = \hat{H}_S+\hat{H}_E+\hat{H}_{\mathrm{Int}} = \frac12
\hat{p}_x^2 + \frac12 \omega^2_0
 \hat{x}^2 + \sum_{n=1}^N \left( \frac12 \hat{p}_{q_n}^2 + \frac12
\omega_n^2 \hat{q}_n^2  + \lambda_n \hat{q}_n \hat{x} \right) \,.
\end{equation}
The perturbative solution to the von Neumann equation
(\ref{vNeumannEquation}) is the Dyson series which is truncated at
second order. We can now trace over the environmental degrees of
freedom to obtain:
\begin{equation}\label{MasterEquation1}
\dot{ \hat{\rho}}_{\mathrm{red}}^{\mathrm{I}}(t) =
\frac{1}{i\hbar} \mathrm{Tr}_E [
\hat{H}_{\mathrm{Int}}^{\mathrm{I}}(t),\hat{\rho}(0)] -
\frac{1}{\hbar^{2}} \int_{0}^{t} \mathrm{d} t_1 \mathrm{Tr}_E [
\hat{H}_{\mathrm{Int}}^{\mathrm{I}}(t), [
\hat{H}_{\mathrm{Int}}^{\mathrm{I}}(t_1),\hat{\rho}(0)]] \,.
\end{equation}
Here, the superscript $\mathrm{I}$ denotes the interaction picture
such that $\hat{H}_{\mathrm{Int}}^{\mathrm{I}}(t) =
\hat{U}_0^{\dag} \hat{H}_{\mathrm{Int}}(0)\hat{U}_0$, with
$\hat{U}_0 = \exp[-i/\hbar(\hat{H}_S+\hat{H}_E)]$. If one now
assumes that initially the total density matrix is not entangled,
i.e.: $\hat{\rho}(0) = \hat{\rho}_S(0)\otimes \hat{\rho}_E(0)$,
one can rewrite equation (\ref{MasterEquation1}) in order to
express it solely in terms of $
\hat{\rho}_{\mathrm{red}}^{\mathrm{I}}(t)$:
\begin{eqnarray}\label{MasterEquation2}
\dot{ \hat{\rho}}_{\mathrm{red}}^{\mathrm{I}}(t) &=&
\frac{1}{i\hbar} \mathrm{Tr}_E [
\hat{H}_{\mathrm{Int}}^{\mathrm{I}}(t),
\hat{\rho}_{\mathrm{red}}^{\mathrm{I}}(t)\otimes \hat{\rho}_E(0) ]
- \frac{1}{\hbar^{2}} \int_{0}^{t} \mathrm{d} t_1 \mathrm{Tr}_E [
\hat{H}_{\mathrm{Int}}^{\mathrm{I}}(t), [
\hat{H}_{\mathrm{Int}}^{\mathrm{I}}(t_1),\hat{\rho}_{\mathrm{red}}^{\mathrm{I}}(t)\otimes
\hat{\rho}_E(0) ]] \\
&& + \frac{1}{\hbar^2} \int_0^{t}\mathrm{d}t_1 \mathrm{Tr}_E
\left( [ \hat{H}_{\mathrm{Int}}^{\mathrm{I}}(t), \mathrm{Tr}_E
\left( [ \hat{H}_{\mathrm{Int}}^{\mathrm{I}}(t_1),
\hat{\rho}_{\mathrm{red}}^{\mathrm{I}}(t)\otimes \hat{\rho}_E(0) ]
\right) \otimes \hat{\rho}_E(0) ]\right)\nonumber \,.
\end{eqnarray}
For the model under consideration (\ref{HamiltonianQM1}), the
master equation above reduces in the Schr\"odinger picture to
($\hbar=1$ again):
\begin{equation}\label{MasterEquation3}
\dot{ \hat{\rho}}_{\mathrm{red}}(t) = \frac{1}{i} [
\hat{H}_{S}(t),\hat{\rho}_{\mathrm{red}}(t) ] - \int_{0}^{t}
\mathrm{d} t_1 \nu(t_1) [\hat{x},[\hat{x}(-t_1),
\hat{\rho}_{\mathrm{red}}(t)]] - i \eta(t_1) [\hat{x}, \{
\hat{x}(-t_1), \hat{\rho}_{\mathrm{red}}(t) \}]\,.
\end{equation}
Here, note that $\hat{x}(-t) = \hat{x}\cos(\omega_0 t) -
\hat{p}_x/\omega_0 \sin(\omega_0 t)$ due to changing back to the
Schr\"odinger picture from the interaction picture. The noise and
dissipation kernels $\nu(t)$ and $\eta(t)$ read at the lowest
order in perturbation theory:
\begin{subequations}
\label{MasterEquation4}
\begin{eqnarray}
\nu(t) &=& \sum_{n=1}^N  \frac{\lambda_n^2 }{2}\langle \{
\hat{q}_n(t), \hat{q}_n(0)\}\rangle = \sum_{n=1}^N \lambda_n^2
\frac{\cos(\omega_n
t)}{2\omega_n}\coth\left(\frac{\beta\omega_n}{2}\right)
\label{MasterEquation4a}\\
\eta(t) &=& \sum_{n=1}^N  \frac{ i \lambda_n^2}{2}\langle
[\hat{q}_n(t), \hat{q}_n(0)]\rangle = \sum_{n=1}^N
\frac{\lambda_n^2}{2}\frac{\sin(\omega_n t)}{\omega_n}
\label{MasterEquation4b} \,.
\end{eqnarray}
\end{subequations}
One thus finds:
\begin{equation}\label{MasterEquation5}
\dot{ \hat{\rho}}_{\mathrm{red}}(t) = -i [ \hat{H}_{S}(t) +
\frac{1}{2} \Omega^2(t) \hat{x}^2 ,\hat{\rho}_{\mathrm{red}}(t) ]
- i \gamma(t) [\hat{x},\{\hat{p}_x,
\hat{\rho}_{\mathrm{red}}(t)\}] - D(t) [ \hat{x},[\hat{x},
\hat{\rho}_{\mathrm{red}}(t)]] - f(t) [\hat{x}, [\hat{p}_x,
\hat{\rho}_{\mathrm{red}}(t)]],
\end{equation}
where the frequency ``renormalisation'' $\Omega(t)$, the damping
coefficient $\gamma(t)$ and the two diffusion coefficients $D(t)$
and $f(t)$ are given by:
\begin{subequations}
\label{MasterEquation5AA}
\begin{eqnarray}
\Omega^2(t) &=& -2 \int_0^{t} \mathrm{d} t_1 \eta(t_1)
\cos(\omega_0 t_1) = \sum_{n=1}^{N} \frac{-\lambda_n^2}{\omega_n
(\omega_0^2 - \omega_n^2)} \left[ \omega_n \left( \cos(\omega_0
t)\cos(\omega_n t) -1 \right) + \omega_0 \sin(\omega_0
t)\sin(\omega_n t)\right] \label{MasterEquation5a}\\
\gamma(t) &=&  \int_0^{t} \mathrm{d} t_1 \eta(t_1)
\frac{\sin(\omega_0 t_1)}{\omega_0} =\sum_{n=1}^{N}
\frac{\lambda_n^2}{2 \omega_0 \omega_n (\omega_0^2 - \omega_n^2)}
\left[ \omega_n \cos(\omega_n t)\sin(\omega_0 t) - \omega_0
\cos(\omega_0
t)\sin(\omega_n t)\right] \label{MasterEquation5b}\\
D(t) &=& \int_0^{t} \mathrm{d} t_1 \nu(t_1) \cos(\omega_0 t_1) =
\sum_{n=1}^{N} \frac{\lambda_n^2
\coth\left(\frac{\beta\omega_n}{2}\right)}{2 \omega_n (\omega_0^2
- \omega_n^2)} \left[ \omega_0 \cos(\omega_n t)\sin(\omega_0 t) -
\omega_n \cos(\omega_0
t)\sin(\omega_n t)\right]  \label{MasterEquation5c}\\
f(t) &=& - \int_0^{t} \mathrm{d} t_1 \nu(t_1) \frac{\sin(\omega_0
t_1)}{\omega_0} = \sum_{n=1}^{N} \frac{\lambda_n^2
\coth\left(\frac{\beta\omega_n}{2}\right) }{2\omega_0 \omega_n
(\omega_0^2 - \omega_n^2)} \left[ \omega_0 \left( \cos(\omega_0
t)\cos(\omega_n t) -1 \right) + \omega_n \sin(\omega_0
t)\sin(\omega_n t)\right] \label{MasterEquation5d} .\phantom{11}
\end{eqnarray}
\end{subequations}
The crucial point is that the non-unitarity present in the master
equation in the terms $\gamma(t)$, $D(t)$ and $f(t)$ in equation
(\ref{MasterEquation5}) brings about the entropy increase. This
entropy increase, in turn, is argued to be a good quantitative
measure of the process of decoherence. Surprisingly, we are not
aware of any direct calculation of the entropy in this approach in
generality. Furthermore, it is important to note that the
effective coupling constant is not just given by $\lambda$ but
rather by:
\begin{equation}\label{effectivelambda}
\lambda_n^{\mathrm{eff}} = \frac{\lambda_n}{\omega_0^2 -
\omega_n^2} \,.
\end{equation}
This can be readily appreciated from examining the
eigenfrequencies for $N=1$ in equation (\ref{EOMOscillator4}) for
small couplings. We will return to this observation in subsequent
sections, but let us at this stage already point out that when the
system and environmental frequencies are close together, one risks
leaving the regime where the master equation is valid. This is
disturbing. When both frequencies are roughly equal one is in the
so-called resonant regime, i.e.: the regime where the interaction
between both oscillators is most effective. There is in principle
nothing non-perturbative about the resonant regime. For example,
one can study the case $N=1$ where $\omega_0 \simeq \omega_1$ and
with both $\lambda/\omega_0^2 \ll 1$ and $\lambda/\omega_1^2 \ll
1$. Despite of this, the coefficients in the master equation tend
to become large in this regime.

Before we solve the master equation (\ref{MasterEquation5}), let
us outline the rather well known ``influence functional''
derivation of the master equation which gives the same
perturbative result \cite{Caldeira:1982iu, Paz:2000le}. The full
density matrix is first projected on position basis:
\begin{equation} \label{CLreducedrho1}
\rho(x,q;y,r,t) = \int \mathrm{d}x' \mathrm{d}y' \mathrm{d}q'
\mathrm{d}r' \langle x, q | e^{-i \hat{H}t} | x', q' \rangle
\langle x', q'| \hat{\rho}(0)| y',r' \rangle \langle y', r' | e^{i
\hat{H}t}| y, r \rangle\,.
\end{equation}
One can then define a quantum mechanical kernel
\cite{Koksma:2007uq} to be the position space projection of the
evolution operator, which can be evaluated using a path integral
approach:
\begin{subequations}
\label{CLreducedrho2}
\begin{eqnarray}
K(x,q,t;x',q',0) &=& \langle x, q | e^{-i \hat{H}t} | x', q'
\rangle = \int \mathcal{D}x\mathcal{D}q \, e^{i S[x,q]}
\label{CLreducedrho2a} \\
K^{\ast}(y,r,t;y',r',0) &=&  \langle y', r' | e^{i \hat{H}t}| y, r
\rangle = \int \mathcal{D}y\mathcal{D}r \, e^{-i S[y,r]}
\label{CLreducedrho2b}\,,
\end{eqnarray}
\end{subequations}
where the action follows from equation (\ref{action:QM4}). The
boundary conditions for the path integrals are given by: $x(t)=x$,
$x(0)=x'$, $q(t)=q$, $q(0)=q'$ and likewise $y(t)=y$, $y(0)=y'$,
$r(t)=r$, $r(0)=r'$. Now, we can trace over the environmental
degrees of freedom as usual to find the reduced density matrix.
Assuming that the system and environment are not entangled
initially, i.e.: $\hat{\rho}(0) = \hat{\rho}_S(0)\otimes
\hat{\rho}_E(0)$, one can thus rearrange the integrals to derive:
\begin{equation} \label{CLreducedrho3}
\rho_{\mathrm{red}}(x,y,t)  = \int \mathrm{d}q \langle x, q
|\hat{\rho}(t)| y, q \rangle = \int \mathrm{d}x' \mathrm{d}y'
J(x,y,t;x',y',0) \rho_S(x',y',0)\,,
\end{equation}
where:
\begin{equation} \label{CLreducedrho4}
J(x,y,t;x',y',0) = \int \mathcal{D}x \mathcal{D}y \, e^{i
S_S[x]-iS_S[y]} \mathcal{F}(x,y)\,,
\end{equation}
and where the influence functional $\mathcal{F}$ is given by:
\begin{equation} \label{CLreducedrho5}
\mathcal{F}(x,y)= \int \mathrm{d}q \mathrm{d}q' \mathrm{d}r'
\rho_E(q',r',0) \int \mathcal{D}q \mathcal{D}r \, e^{i
S_E[q]-iS_E[r]+ i S_{\mathrm{Int}}[x,q]-iS_{\mathrm{Int}}[y,r] }
\,.
\end{equation}
Caldeira and Leggett then derive an equation for
\mbox{$\dot{\rho}_{\mathrm{red}}(x,y,t)$}, i.e. the master
equation, by carefully studying infinitesimal time translations
\mbox{$\rho_{\mathrm{red}}(x,y,t+\epsilon)$}, which they expand in
$\epsilon$.

\section{Two Coupled Oscillators}
\label{Two Coupled Oscillators}

We can specialise to the case of two bilinearly coupled simple
harmonic oscillators by setting $N=1$ in
equation~(\ref{action:QM3}). The total Lagrangian we consider in
this section is given by:
\begin{equation}\label{action:QM4}
L\big[x, q \big] = \frac12 {\dot x}^2 - \frac12 \omega^2_0
 x^2 + \frac12 {\dot q}^2 - \frac12
\omega_1^2 q^2  - \lambda q x \,.
\end{equation}
This is the simplest quantum mechanical system one can think of
that captures some of the physics that is relevant for
decoherence. In this section, we perform two computations:
\begin{itemize}
\item Decoherence in the conventional approach: we can straightforwardly solve for the master
equation in this case;
\item Decoherence in our correlator approach: we can solve for the statistical propagator of the system
oscillator and the environment oscillator by exact analytic
methods from which we can extract the exact Gaussian entropy.
\end{itemize}
The bottom line is clear: the simple case (\ref{action:QM4})
allows for a clean comparison of the two approaches to
decoherence.

In appendix \ref{Reducing the Density Matrix} we compute the
reduced density matrix by tracing over the environmental degrees
of freedom exactly as the Lagrangian (\ref{action:QM4}) is
Gaussian. As we can also solve the von Neumann equation for the
full density matrix, we can study the (unitary) time evolution of
the reduced density matrix and confirm that the equality derived
in equation (\ref{entropyequality}) is valid.

\subsection{The Master Equation Approach to Decoherence}
\label{The Master Equation Approach to Decoherence}

We are now ready to evaluate the master equation
(\ref{MasterEquation5}). Let us project this operator equation on
the position bras and kets as follows\footnote{Alternatively one
could calculate $\rho_{\mathrm{red}}$ using the well known
Gaussian propagator of the reduced theory \cite{Calzetta:book}.}:
\begin{equation} \label{reduceddensitymatrixAnsatz}
\rho_{\mathrm{red}}(x,y;t) = \langle x|
\hat{\rho}_{\mathrm{red}}(t)| y \rangle=
\tilde{\mathcal{N}}(t)\exp\left[
-\tilde{a}(t)x^2-\tilde{a}^{\ast}(t)y^2+2\tilde{c}(t)xy \right]\,.
\end{equation}
The master equation (\ref{MasterEquation5}) thus reduces to the
following coupled system of differential equations:
\begin{subequations}
\label{MasterEquation6}
\begin{eqnarray}
\frac{\mathrm{d} \tilde{a}_{\mathrm{R}}}{\mathrm{d}t} &=& 4
\tilde{a}_{\mathrm{I}} \tilde{a}_{\mathrm{R}} - 2\gamma(t)
(\tilde{a}_{\mathrm{R}} + \tilde{c}) + D(t) - 2 f(t)
\tilde{a}_{\mathrm{I}}
\label{MasterEquation6a} \\
\frac{\mathrm{d}\tilde{a}_{\mathrm{I}}}{\mathrm{d}t} &=& 2\left(
\tilde{a}_{\mathrm{I}}^2 - \tilde{a}_{\mathrm{R}}^2 + \tilde{c}^2
\right) + \frac{1}{2}\left( \omega_0^2+\Omega^2(t)\right) - 2
\gamma(t)
\tilde{a}_{\mathrm{I}} + 2 f(t) (\tilde{a}_{\mathrm{R}} - \tilde{c}) \label{MasterEquation6b} \\
\frac{\mathrm{d}\tilde{c}}{\mathrm{d}t} &=& 4
\tilde{a}_{\mathrm{I}} \tilde{c} - 2 \gamma(t)
(\tilde{a}_{\mathrm{R}} + \tilde{c})+D(t) - 2 f(t)
\tilde{a}_{\mathrm{I}}\label{MasterEquation6c} \\
\frac{\mathrm{d}}{\mathrm{d}t} \ln({\cal \tilde{N}}) &=& 2
\tilde{a}_{\mathrm{I}}\label{MasterEquation6d} \,.
\end{eqnarray}
\end{subequations}
Here, the subscript $I$ denotes the imaginary part
$a_{\mathrm{I}}=\Im[a]$. It turns out to be advantageous to
directly compute the time evolution of our three non-trivial
Gaussian correlators. Analogous methods have been used in
\cite{Blume-Kohout:2003} to analyse decoherence in an upside down
simple harmonic oscillator. Let us start by recalling equation
(\ref{correlators:all}). It is valid for the coefficients of the
reduced density matrix too. We can thus derive the following set
of differential equations:
\begin{subequations}
\label{MasterEquation6Corr}
\begin{eqnarray}
\frac{\mathrm{d}\langle \hat{x}^2\rangle }{\mathrm{d}t} &=& -
\frac{ \dot{\tilde{a}}_{\mathrm{R}} -
\dot{\tilde{c}}}{4(\tilde{a}_{\mathrm{R}} - \tilde{c})^2} = 2
\left \langle \frac{1}{2} \{\hat{x},\hat{p}\}\right \rangle
\label{MasterEquation6Corra} \\
\frac{\mathrm{d}\langle \hat{p}^2\rangle }{\mathrm{d}t} &=& -
2(\omega_0^2+\Omega^2)\left \langle \frac{1}{2}
\{\hat{x},\hat{p}\}\right \rangle - 4 \gamma(t) \langle
\hat{p}^2\rangle + 2D(t) \label{MasterEquation6Corrb} \\
\frac{\mathrm{d}\left \langle \frac{1}{2}
\{\hat{x},\hat{p}\}\right \rangle }{\mathrm{d}t} &=& -
(\omega_0^2+\Omega^2)\langle \hat{x}^2 \rangle + \langle
\hat{p}^2\rangle- f(t) -2 \gamma(t) \left \langle \frac{1}{2}
\{\hat{x},\hat{p}\} \right \rangle \label{MasterEquation6Corrc}
\,.
\end{eqnarray}
\end{subequations}
Initially, we impose that the system is in a pure state:
\begin{subequations}
\label{MasterEquation7}
\begin{eqnarray}
\langle \hat{x}^2(t_0)\rangle &=& \frac{1}{2\omega_0}
\label{MasterEquation7a} \\
\langle \hat{p}^2 (t_0) \rangle  &=& \frac{\omega_0}{2}
\label{MasterEquation7b} \\
\left \langle \frac{1}{2} \{\hat{x},\hat{p}\}\right \rangle &=& 0
\,. \label{MasterEquation7c}
\end{eqnarray}
\end{subequations}
We can then straightforwardly find the phase space area in
equation (\ref{PhaseSpaceArea1}) and the von Neumann entropy for
the system now follows via equation (\ref{vNeumannEntropy2}). We
discuss the results after having solved for the statistical
propagator in the correlator approach to decoherence in the next
subsection.

\subsection{The Correlator Approach to Decoherence}
\label{The Correlator Approach to Decoherence}

The statistical propagator in quantum mechanics contains all the
relevant information about the three non-trivial Gaussian
correlators characterising our system as is apparent from equation
(\ref{PhaseSpaceArea1}). Rather than separately solving for these
three correlators by means of calculating various expectation
values, one can directly solve for the statistical propagator.
This method is analogous in spirit to the quantum field
theoretical computations we performed in \cite{Koksma:2009wa},
though there we solved the Kadanoff-Baym equations for the system
field $\phi$ to obtain the correlators\footnote{Indeed such
correlator equations also follow on the way to the master equation
if one inverts the kernel in the Gaussian exponential of the path
integral for $J$ in equation (\ref{CLreducedrho4}). The resulting
equations are 1PI. Usually, they are obtained by a double Legendre
transform \cite{Cornwall:1974vz, Jackiw:1974cv, Koksma:2009wa}.}.
The oscillator model of interest is admittedly very simple, so one
could use many equivalent methods to solve for the correlators.
The point of our calculation is that in cases of real physical
interest, such as the one examined in \cite{Koksma:2009wa}, it is
the only tractable method to perform the calculation since for
interacting (non-Gaussian) models it is very hard to solve for the
total density matrix, even perturbatively. It is possible however
to perturbatively solve for the statistical propagator.

As the derivation of the statistical propagator for general
initial conditions, i.e.: initial conditions that also permit
entangled states, is straightforward but rather cumbersome, we
present a full derivation in appendix \ref{The Statistical
Propagator for N1}. The principle on which this derivation relies
is to diagonalise the coupled equations of motion for the two
oscillators such that they decouple. One can then solve the
differential equations trivially, rotate back to the original
frame and impose initial conditions. It is important to realize
that our expression for the statistical propagator in equation
(\ref{statisticalpropagatorsolution2}) is exact.

\subsubsection{Initial Conditions I: Pure and Thermal Initial
States}

In particular, we are interested in what we henceforth call
``pure-thermal'' initial conditions. Here, pure refers to the
initial state of the system, whereas thermal refers to the initial
state of the environment field. Since we are dealing with only one
environmental simple harmonic oscillator, the phrase ``thermal
initial condition'' might be a little awkward. We only intend to
imply that the environmental statistical propagator probes one
point on Planck's distribution. We need of course many more
oscillators to sensibly talk about a thermal distribution of
oscillators. Also, we require that the system and environment are
separable initially, such that no system-environment correlations
are present at $t_0$. Using equation
(\ref{propagatorsstatistical}), we thus require:
\begin{equation} \label{purethermal_initialconditions}
\begin{array}{lllllll}
\langle \hat{x}^2(t_0)\rangle = \frac{1}{2\omega_0} & \phantom{1}
& \langle \hat{p}_x^2(t_0)\rangle = \frac{\omega_0}{2} &
\phantom{1} & \langle \{ \hat{x}(t_0),\hat{p}_x(t_0) \}\rangle  =
0 & \phantom{1} & \phantom{1} \\
\langle \hat{q}^2(t_0)\rangle =
\frac{1}{2\omega_1}\coth\left(\frac{\beta \omega_1}{2}\right) &
\phantom{1} & \langle \hat{p}_q^2(t_0)\rangle =
\frac{\omega_1}{2}\coth\left(\frac{\beta \omega_1}{2}\right) &
\phantom{1} & \langle \{ \hat{q}(t_0),\hat{p}_q(t_0) \}\rangle = 0
& \phantom{1}& \phantom{1} \\
\langle \hat{x}(t_0) \hat{q}(t_0)\rangle =0 &  \phantom{1} &
\langle \hat{p}_x(t_0) \hat{p}_q(t_0) \rangle = 0 & \phantom{1} &
\langle \hat{x}(t_0)\hat{p}_q(t_0) \rangle = 0 & \phantom{1}&
\langle \hat{q}(t_0)\hat{p}_x(t_0) \rangle = 0\,,
\end{array}
\end{equation}
Having the statistical propagator in equation
(\ref{statisticalpropagatorsolution2}) at our disposal, we can now
find the phase space area and the resulting entropy in various
interesting cases.

\subsubsection{Initial Conditions II: Time Translation Invariant States}
\label{Initial Conditions II: Time Translation Invariant States}

It turns out there is a specific class of initial conditions that,
as we shall see, do not give rise to oscillatory behaviour in e.g.
the entropy. Expectation values in this case are a constant
despite of the coupling $\lambda$ between the two oscillators. We
present a derivation at the end of appendix \ref{The Statistical
Propagator for N1}. Initially, we require:
\begin{equation} \label{timetranslationinvariantstate_initialconditions}
\begin{array}{lllllll}
\langle \hat{x}^2(t_0)\rangle = \frac{1}{2\omega_0} & \phantom{1}
& \langle \hat{p}_x^2(t_0)\rangle = \frac{\omega_0}{2} &
\phantom{1} & \langle \{ \hat{x}(t_0),\hat{p}_x(t_0) \}\rangle  =
0 & \phantom{1} & \phantom{1} \\
\langle \hat{q}^2(t_0)\rangle = \frac{1}{2\omega_0}& \phantom{1} &
\langle \hat{p}_q^2(t_0)\rangle = \frac{\omega_1^2}{2\omega_0}&
\phantom{1} & \langle \{ \hat{q}(t_0),\hat{p}_q(t_0) \}\rangle = 0
& \phantom{1}& \phantom{1} \\
\langle \hat{x}(t_0) \hat{q}(t_0)\rangle =0 &  \phantom{1} &
\langle \hat{p}_x(t_0) \hat{p}_q(t_0) \rangle =
\frac{\lambda}{2\omega_0} & \phantom{1} & \langle
\hat{x}(t_0)\hat{p}_q(t_0) \rangle = 0 & \phantom{1}& \langle
\hat{q}(t_0)\hat{p}_x(t_0) \rangle = 0\,,
\end{array}
\end{equation}
Clearly, the system is in a pure state and the environmental
oscillator is in a ``thermal'' state whose temperature is fixed by
requiring:
\begin{equation}\label{thermalcondition}
\coth\left( \frac{\beta \omega_1}{2}\right) =
\frac{\omega_1}{\omega_0} \,.
\end{equation}
Finally, we see that the initial state is entangled, i.e.: the
initial state is \emph{not} of the form $\hat{\rho}(0) =
\hat{\rho}_S(0)\otimes \hat{\rho}_E(0)$.

\subsection{Results}
\label{Results}

%
\begin{figure}
    \begin{minipage}[t]{.45\textwidth}
        \begin{center}
\includegraphics[width=\textwidth]{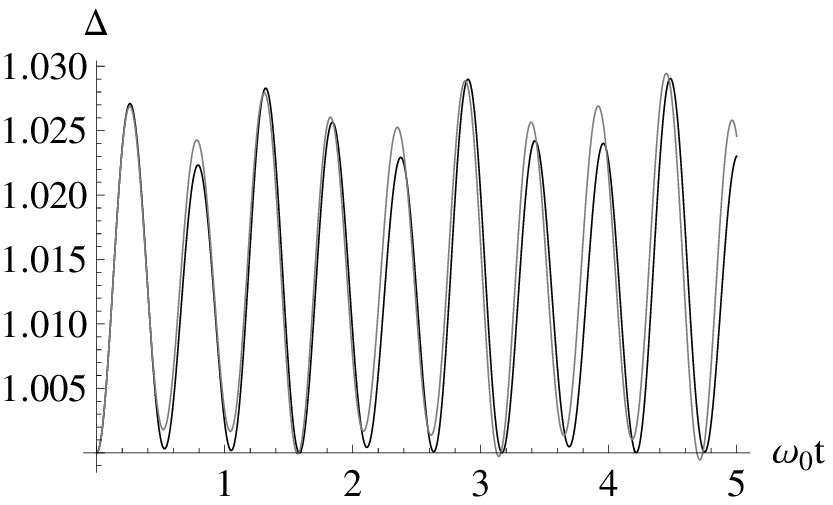}
   {\em \caption{Phase space area as a function of time. Solid black: correlator approach to decoherence.
   Solid gray: master equation approach to decoherence. We use $\omega_1/\omega_0=2$, $\lambda/\omega_0^2=1/2$ and $\beta\omega_0=2000$.
   \label{fig:Delta_T0} }}
        \end{center}
   \end{minipage}
\hfill
    \begin{minipage}[t]{.45\textwidth}
        \begin{center}
\includegraphics[width=\textwidth]{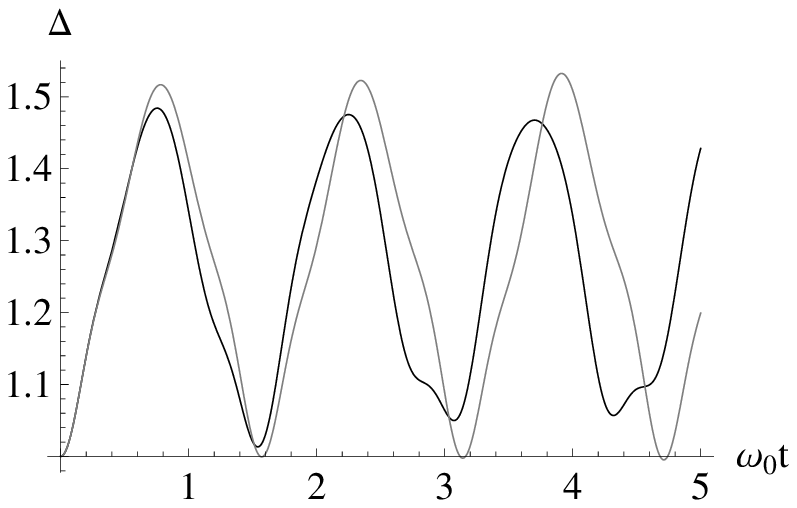}
   {\em \caption{Phase space area as a function of time. Solid black: correlator approach to decoherence.
   Solid gray: master equation approach to decoherence. We use $\omega_1/\omega_0=2$, $\lambda/\omega_0^2=1/2$ and $\beta\omega_0=0.2$.
    \label{fig:Delta_T} }}
        \end{center}
    \end{minipage}
\vskip 0.1cm
    \begin{minipage}[t]{.45\textwidth}
        \begin{center}
\includegraphics[width=\textwidth]{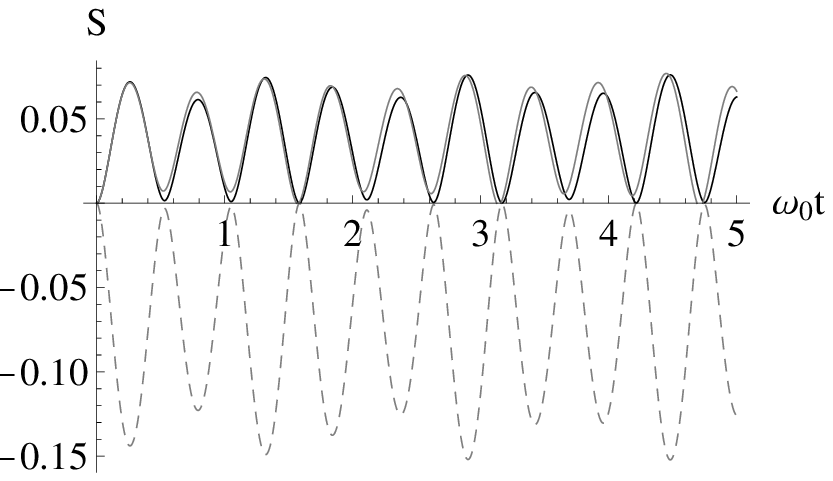}
   {\em \caption{Entropy as a function of time. Solid black: correlator approach to decoherence, which follows from equation (\ref{vNeumannEntropy6}).
   Solid gray: master equation approach to decoherence. Dashed gray: correlation entropy, which follows from equation (\ref{vNeumannEntropy7}).
   We use $\omega_1/\omega_0=2$, $\lambda/\omega_0^2=1/2$ and $\beta\omega_0=2000$.
   \label{fig:S_T0} }}
        \end{center}
    \end{minipage}
\hfill
    \begin{minipage}[t]{.45\textwidth}
        \begin{center}
\includegraphics[width=\textwidth]{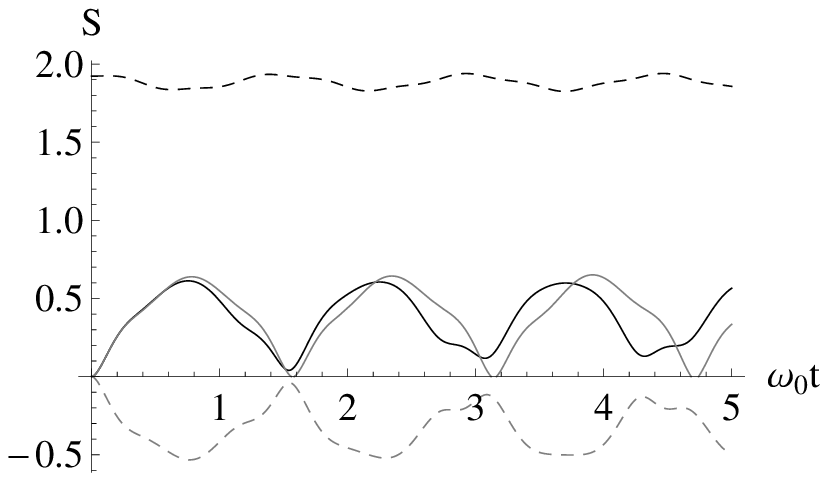}
   {\em \caption{Entropy as a function of time. Solid black: correlator approach to decoherence.
   Solid gray: master equation approach to decoherence. Dashed black: entropy of the environment oscillator.
   Dashed gray: correlation entropy. We use $\omega_1/\omega_0=2$, $\lambda/\omega_0^2=1/2$ and $\beta\omega_0=0.2$.
   \label{fig:S_T} }}
        \end{center}
    \end{minipage}
\vskip 0.1cm
    \begin{minipage}[t]{.45\textwidth}
        \begin{center}
\includegraphics[width=\textwidth]{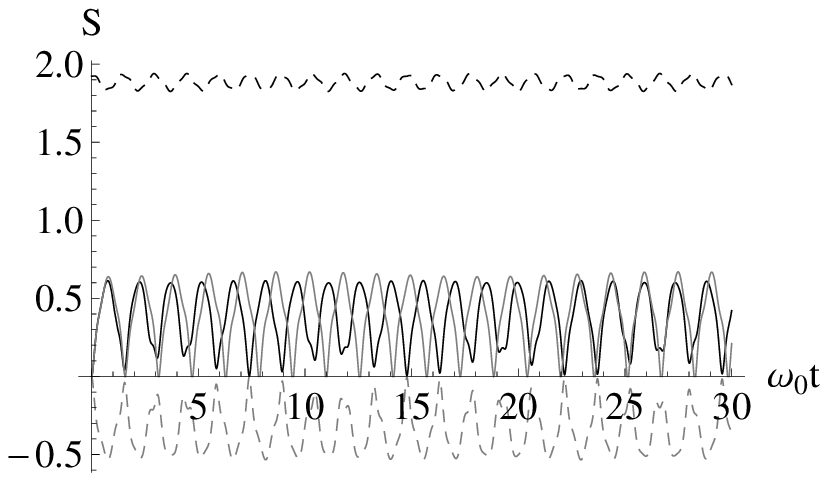}
   {\em \caption{Entropy as a function of time. Solid black: correlator approach to decoherence.
   Solid gray: master equation approach to decoherence. Dashed black: entropy of the environment oscillator.
   Dashed gray: correlation entropy. We use $\omega_1/\omega_0=2$, $\lambda/\omega_0^2=1/2$ and $\beta\omega_0=0.2$. The quasi periodic
   behaviour is due to
   Poincar\'e's recurrence theorem.
    \label{fig:S_Tlatetimes}}}
        \end{center}
    \end{minipage}
\hfill
    \begin{minipage}[t]{.45\textwidth}
        \begin{center}
\includegraphics[width=\textwidth]{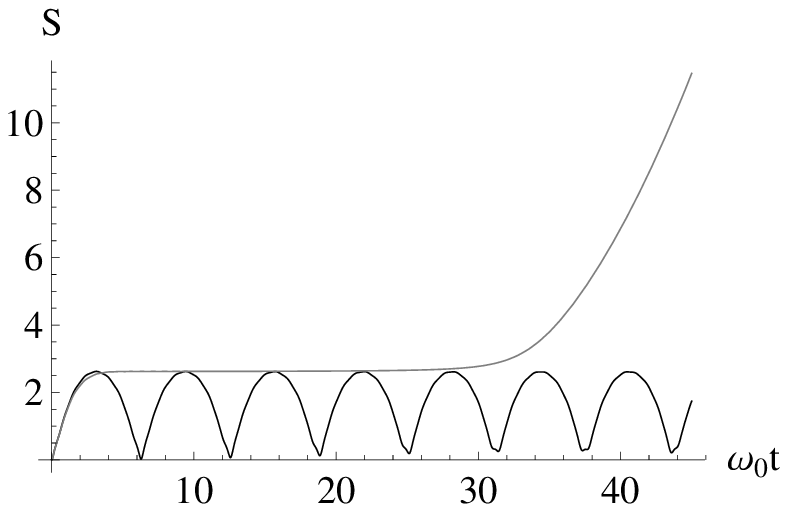}
   {\em \caption{Entropy as a function of time. Solid black: correlator approach to decoherence.
   Solid gray: master equation approach to decoherence. We use $\omega_1/\omega_0=201/200$, $\lambda/\omega_0^2=1/4$ and $\beta\omega_0=0.2$ such that we
   are in the resonant regime. The entropy as obtained from the
   master equation shows unphysical behaviour.
   \label{fig:Sresregime} }}
        \end{center}
    \end{minipage}
\end{figure}

Let us begin by examining figures~\ref{fig:Delta_T0}
and~\ref{fig:Delta_T}. In these figures we show the phase space
area as a function of time for both the master equation approach
to decoherence (gray) and our correlator approach to decoherence
(black). In figure~\ref{fig:Delta_T0} the environment is,
effectively, at $T=0$ as we use $\beta\omega_0=2000$. In
figure~\ref{fig:Delta_T} we use $\beta\omega_0=0.2$ such that the
environmental oscillator is in ``thermal equilibrium'' at some
finite temperature $k_B T = 5 \omega_0$. The difference between
the two phase space areas is perturbatively small in both cases
(the effective frequency that enters the master equation has been
calculated by perturbative methods). In other words: both methods
agree well. We can thus conclude that the entropy one obtains from
the perturbative master equation in the Gaussian case in fact
stems from neglecting the correlation entropy, just as in our
correlator approach to decoherence.

As energy is conserved in our model, the Poincar\'e recurrence
theorem applies. This theorem states that our system will after a
sufficiently long time return to a state arbitrary close to its
initial state. The Poincar\'e recurrence time is the amount of
time this takes. In figure~\ref{fig:Delta_T0} we used almost
commensurate eigenfrequencies. Consequently, the behaviour of the
phase space area as a function of time is very regular.
Introducing a non-zero temperature in figure~\ref{fig:Delta_T}
automatically makes the eigenfrequencies non-commensurate and thus
induces a greater irregularity in the time evolution of the phase
space area. In other words: the Poincar\'e recurrence time has
increased.

Let us recall equation (\ref{vNeumannEntropy5}):
\begin{equation}
S_{\mathrm{vN}} = S_{\mathrm{total}} = S_{S}(t)+
S_{E}(t)+S_{SE}(t)\nonumber \,.
\end{equation}
In figures~\ref{fig:S_T0},~\ref{fig:S_T}
and~\ref{fig:S_Tlatetimes} we show the Gaussian von Neumann
entropy for the system $S_{S}(t)$ (solid black) and for the
environment $S_{E}(t)$ (dashed black), the Gaussian correlation
entropy $S_{SE}(t)$ (dashed gray) and, finally, the entropy
resulting from the master equation
$S_{\mathrm{vN}}^{\mathrm{red}}(t)$ (solid gray). The solid black
line in figure~\ref{fig:S_T0} is the entropy $S_{S}(t)$ one
obtains from figure~\ref{fig:Delta_T0}. Likewise, the solid black
lines in figures~\ref{fig:S_T} and~\ref{fig:S_Tlatetimes} show the
entropy $S_{S}(t)$ as a function of time one gets from
figure~\ref{fig:Delta_T}. In figure~\ref{fig:S_Tlatetimes} we show
the behaviour up to very large times to nicely illustrate the
quasi periodicity resulting from Poincar\'e's recurrence theorem.

The dashed black lines in figures~\ref{fig:S_T}
and~\ref{fig:S_Tlatetimes} show the environmental entropy
$S_{E}(t)$. Just as $S_S (t)$, one can obtain $S_{E}(t)$ from the
three non-trivial Gaussian correlators $\langle \hat{q}^2
\rangle$, $\langle \hat{p}_q^2 \rangle$ and $1/2\langle\{\hat
q,\hat p_q\}\rangle$ and by making use of equation
(\ref{PhaseSpaceArea1}). The environmental entropy in
figure~\ref{fig:S_T0} is precisely equal to the system entropy
$S_S (t)$ because both the system and the environment are at
$T=0$. The dashed gray lines in
figures~\ref{fig:S_T0},~\ref{fig:S_T} and~\ref{fig:S_Tlatetimes}
show the Gaussian correlation entropy $S_{SE}(t)$ one obtains from
equation (\ref{vNeumannEntropy7}). Finally, we can easily compare
with the reduced von Neumann entropy one obtains from solving the
master equation (in solid gray). Just as for the phase space area,
the results differ only due to a small perturbative error in the
master equation.

The oscillatory behaviour in the entropy can be interpreted in two
ways:
\begin{itemize}
\item If one averages out the oscillations, the entropy of the
system oscillator evolves from $S_S (0)=0$ to some small non-zero
value $\overline{S}_S >0$, where $\overline{S}_S $ is the
(classical) time average $\overline{S}_S = \langle S_S(t)
\rangle$. This average non-zero value of the system entropy
$\overline{S}_S $ exists by virtue of a negative correlation
entropy. The amount of decoherence that the system has experienced
thus equals $\overline{S}_S $;
\item The Poincar\'e recurrence theorem enforces that the system
returns arbitrarily close to its initial state after some finite
Poincar\'e recurrence time. In quantum mechanics this recurrence
time is rather small. This is not the case in field theoretical
models as we discuss shortly.
\end{itemize}

So far, it is clear that the phase space area or the entropy in
our correlator approach to decoherence and in the master equation
approach to decoherence agree well up to perturbative corrections.
Let us now examine the so-called resonant regime, where $\omega_1
\simeq \omega_0$ in figure \ref{fig:Sresregime}. Here, we set
$\omega_1/\omega_0=201/200$. Our coupling is still small in this
case so we are well in the perturbative regime where the master
equation should yield sensible behaviour. Clearly, it does not as
the entropy blows up. In the resonant regime the master equation
suffers from secular growth which physically is not acceptable.
Our approach indeed yields a perfectly finite evolution for the
entropy.

\begin{figure}
     \begin{minipage}[t]{.45\textwidth}
        \begin{center}
\includegraphics[width=\textwidth]{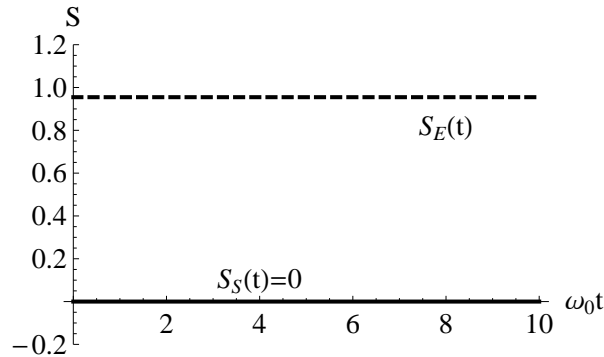}
   {\em \caption{Entropy as a function of time. Solid black: entropy of the system. Dashed black: entropy of the environment oscillator.
   We use $\omega_1/\omega_0=2$, $\lambda/\omega_0^2=1/4$ and $\beta$ is thus fixed via relation (\ref{thermalcondition}).
   The system entropy vanishes (pure state) and is time
   independent. The environmental entropy is a
   constant too. Despite that $\lambda \neq 0$, there are no
   oscillations in the expectation values as shown above.
   \label{fig:Stimeindep} }}
        \end{center}
    \end{minipage}
\end{figure}
In figure~\ref{fig:Stimeindep} we show the time evolution of the
entropy for the time translation invariant initial conditions in
equation (\ref{timetranslationinvariantstate_initialconditions})
in our correlator approach to decoherence. Neither the entropy nor
the phase space area change in time. We find this result rather
interesting. Despite the fact that we are dealing with a non-zero
coupling $\lambda \neq 0$, we can fine tune our initial conditions
(\ref{timetranslationinvariantstate_initialconditions}) such that
there appears to be no influence of the coupling whatsoever when
the entropy is measured. In other words, the expectation values in
equation (\ref{timetranslationinvariantstate_initialconditions})
are time independent and consequently fixed by their initial value
despite of the coupling. This does not depend on how large the
coupling $\lambda$ is.

\section{N Coupled Oscillators}
\label{N Coupled Oscillators}

We are really interested in the general case of $N$-oscillators.
Let us consider the original Lagrangian in equation
(\ref{action:QM}). We take $N=50$ throughout this section. In this
section, we perform again two calculations:
\begin{itemize}
\item Decoherence in the conventional approach: we can numerically solve for the master
equation in this case;
\item Decoherence in our correlator approach: we can solve for the three non-trivial
Gaussian system correlators by exact numerical methods from which
we calculate the entropy.
\end{itemize}
Again does our simple action in equation (\ref{action:QM}) allow
for a clean comparison of the two approaches to decoherence.

\subsection{The Master Equation Approach to Decoherence}
\label{The Master Equation Approach to Decoherence2}

The form of the Paz-Zurek master equation itself remains unchanged
of course and is given in equation (\ref{MasterEquation5}) and
(\ref{MasterEquation5AA}). Compared to the previous section where
we studied two oscillators only the precise form of the
coefficients in equation (\ref{MasterEquation5AA}) changes.
Therefore, we exploit the same strategy as in the previous section
and solve for the three non-trivial Gaussian correlators as in
equation (\ref{MasterEquation6Corr}).

\subsection{The Correlator Approach to Decoherence}
\label{The Correlator Approach to Decoherence2}

The simplest way to study $N=50$ oscillators is not to diagonalise
the resulting equation of motion (although this is of course
possible in principle), but just study the unitary evolution of
all Gaussian correlators in our system of interest by numerical
methods. Let us recall the total Hamiltonian of our system of
interest given in equation (\ref{HamiltonianQM1}):
\begin{equation}
\hat{H} = \hat{H}_S+\hat{H}_E+\hat{H}_{\mathrm{Int}} = \frac12
\hat{p}_x^2 + \frac12 \omega^2_0
 \hat{x}^2 + \sum_{n=1}^N \left( \frac12 \hat{p}_{q_n}^2 + \frac12
\omega_n^2 \hat{q}_n^2  + \lambda_n \hat{q}_n \hat{x} \right)
\nonumber \,.
\end{equation}
Hamilton's equations of motion thus yield:
\begin{subequations}
\label{HamiltonianQM2}
\begin{eqnarray}
\dot{\hat{x}} &=& \hat{p}_x \label{HamiltonianQM2a} \\
\dot{\hat{p}}_x &=& -\omega_0^2 \hat{x} - \sum_{n=1}^{N} \lambda_n \hat{q}_n \label{HamiltonianQM2b} \\
\dot{\hat{q}}_n &=& \hat{p}_{q_n}  \label{HamiltonianQM2c} \\
\dot{\hat{p}}_{q_n} &=& -\omega_n^2 \hat{q}_n - \lambda_n \hat{x}
\label{HamiltonianQM2d}  \,.
\end{eqnarray}
\end{subequations}
We can thus straightforwardly derive the following set of coupled
linear first order differential equations that govern the unitary
time evolution of all Gaussian correlators:
\begin{subequations}
\label{HamiltonianQM3}
\begin{eqnarray}
\frac{\mathrm{d}\langle \hat{x}^2\rangle }{\mathrm{d}t} &=& 2
\left \langle \frac{1}{2} \{\hat{x},\hat{p}_x\}\right \rangle
\label{HamiltonianQM3a} \\
\frac{\mathrm{d}\langle \hat{p}_x^2\rangle }{\mathrm{d}t} &=& - 2
\omega_0^2 \left \langle \frac{1}{2} \{\hat{x},\hat{p}_x\}\right
\rangle -2 \sum_{n=1}^{N} \lambda_n \langle \hat{p}_x \hat{q}_n  \rangle \label{HamiltonianQM3b} \\
\frac{\mathrm{d}\left \langle \frac{1}{2}
\{\hat{x},\hat{p}_x\}\right \rangle }{\mathrm{d}t} &=& -
\omega_0^2\langle \hat{x}^2 \rangle + \langle \hat{p}_x^2\rangle -
\sum_{n=1}^{N} \lambda_n \langle \hat{x}\hat{q}_n\rangle
\label{HamiltonianQM3c}\\
\frac{\mathrm{d}\langle \hat{q}^2_n\rangle }{\mathrm{d}t} &=& 2
\left \langle \frac{1}{2} \{\hat{q}_n,\hat{p}_{q_n}\}\right
\rangle
\label{HamiltonianQM3d} \\
\frac{\mathrm{d}\langle \hat{p}^2_{q_n}\rangle }{\mathrm{d}t} &=&
- 2 \omega_n^2 \left \langle \frac{1}{2}
\{\hat{q}_n,\hat{p}_{q_n}\}\right
\rangle -2 \lambda_n \langle \hat{p}_{q_n} \hat{x}\rangle   \label{HamiltonianQM3e} \\
\frac{\mathrm{d}\left \langle \frac{1}{2}
\{\hat{q}_n,\hat{p}_{q_n}\}\right \rangle }{\mathrm{d}t} &=& -
\omega_n^2\langle \hat{q}^2_n \rangle + \langle
\hat{p}^2_{q_n}\rangle - \lambda_n \langle \hat{x}\hat{q}_n\rangle
\label{HamiltonianQM3f} \\
\frac{\mathrm{d}\langle \hat{x}
\hat{q}_n\rangle }{\mathrm{d}t} &=& \langle
\hat{p}_x\hat{q}_n\rangle +  \langle \hat{x}\hat{p}_{q_n}\rangle
\label{HamiltonianQM3g} \\
\frac{\mathrm{d}\langle \hat{p}_x \hat{q}_n\rangle }{\mathrm{d}t}
&=& \langle \hat{p}_x\hat{p}_{q_n}\rangle - \omega_0^2 \langle
\hat{x} \hat{q}_n\rangle - \sum_{i=1}^{N}\lambda_i \langle
\hat{q}_i \hat{q}_n\rangle
 \label{HamiltonianQM3h} \\
\frac{\mathrm{d}\langle \hat{x} \hat{p}_{q_n}\rangle
}{\mathrm{d}t} &=& \langle \hat{p}_x\hat{p}_{q_n}\rangle -
\omega_n^2 \langle \hat{x} \hat{q}_n\rangle -\lambda_n \langle
\hat{x}^2\rangle \label{HamiltonianQM3i}\\
\frac{\mathrm{d}\langle \hat{p}_x \hat{p}_{q_n}\rangle
}{\mathrm{d}t} &=& -\omega_0^2 \langle \hat{x}\hat{p}_{q_n}
\rangle - \omega_n^2 \langle \hat{p}_x \hat{q}_n\rangle -\lambda_n
\left\langle \frac{1}{2} \{ \hat{x}, \hat{p}_x\}\right \rangle
-\lambda_n \left\langle \frac{1}{2} \{ \hat{q}_n, \hat{p}_{q_n}
\}\right \rangle - \sum_{\substack{i=1 \\
i \neq n}}^{N} \lambda_i \langle \hat{q}_i \hat{p}_{q_n}\rangle
\label{HamiltonianQM3j} \\
\frac{\mathrm{d}\langle \hat{q}_n \hat{q}_m \rangle }{\mathrm{d}t}
&=& \langle \hat{q}_n \hat{p}_{q_m}\rangle +  \langle \hat{q}_m \hat{p}_{q_n}\rangle  \label{HamiltonianQM3k}\\
\frac{\mathrm{d}\langle \hat{q}_n \hat{p}_{q_m}\rangle
}{\mathrm{d}t} &=& \langle \hat{p}_{q_n}\hat{p}_{q_m}\rangle -
\omega_m^2 \langle \hat{q}_n \hat{q}_m\rangle -\lambda_m \langle
\hat{x} \hat{q}_n \rangle \label{HamiltonianQM3l}\\
\frac{\mathrm{d}\langle \hat{p}_{q_n} \hat{p}_{q_m}\rangle
}{\mathrm{d}t} &=& - \omega_n^2 \langle \hat{q}_n
\hat{p}_{q_m}\rangle  - \omega_m^2 \langle \hat{q}_m
\hat{p}_{q_n}\rangle - \lambda_n \langle \hat{x}
\hat{p}_{q_m}\rangle  - \lambda_m \langle \hat{x}
\hat{p}_{q_n}\rangle \label{HamiltonianQM3m}\,.
\end{eqnarray}
\end{subequations}
In the last three equations $m \neq n$ is implied. In other words,
for $m=n$, we would find equations (\ref{HamiltonianQM3d}),
(\ref{HamiltonianQM3e}) and (\ref{HamiltonianQM3f}) again. As in
the previous section, we require ``pure-thermal'' initial
conditions. They generalise to:
\begin{equation} \label{purethermal_initialconditionsN}
\begin{array}{lllllll}
\langle \hat{x}^2(t_0)\rangle = \frac{1}{2\omega_0} & \phantom{1}
& \langle \hat{p}_x^2(t_0)\rangle = \frac{\omega_0}{2} &
\phantom{1} & \langle \{ \hat{x}(t_0),\hat{p}_x(t_0) \}\rangle  =
0 & \phantom{1} & \phantom{1} \\
\langle \hat{q}^2_n(t_0)\rangle =
\frac{1}{2\omega_n}\coth\left(\frac{\beta \omega_n}{2}\right) &
\phantom{1} & \langle \hat{p}_{q_n}^2(t_0)\rangle =
\frac{\omega_n}{2}\coth\left(\frac{\beta \omega_n}{2}\right) &
\phantom{1} & \langle \{ \hat{q}_n(t_0),\hat{p}_{q_n}(t_0)
\}\rangle = 0
& \phantom{1}& \phantom{1} \\
\langle \hat{x}(t_0) \hat{q}_n(t_0)\rangle =0 &  \phantom{1} &
\langle \hat{p}_x(t_0) \hat{p}_{q_n}(t_0) \rangle = 0 &
\phantom{1} & \langle \hat{x}(t_0)\hat{p}_{q_n}(t_0) \rangle = 0 &
\phantom{1}& \langle \hat{q}_n(t_0)\hat{p}_x(t_0) \rangle = 0\\
\langle \hat{q}_n(t_0) \hat{q}_m(t_0)\rangle = 0 & \phantom{1} &
\langle \hat{p}_{q_n}(t_0)\hat{p}_{q_m}(t_0)\rangle = 0&
\phantom{1} & \langle \hat{q}_n(t_0) \hat{p}_{q_m}(t_0) \rangle =
0 \,,& \phantom{1}& \phantom{1}
\end{array}
\end{equation}
where of course $1 \leq \{n,m\} \leq N$, $m \neq n$. For
simplicity we will restrict ourselves to the case where $\lambda =
\lambda_n$, $1 \leq n \leq N$, throughout the paper.

\subsection{Results}
\label{Results2}

%
\begin{figure}
    \begin{minipage}[t]{.45\textwidth}
        \begin{center}
\includegraphics[width=\textwidth]{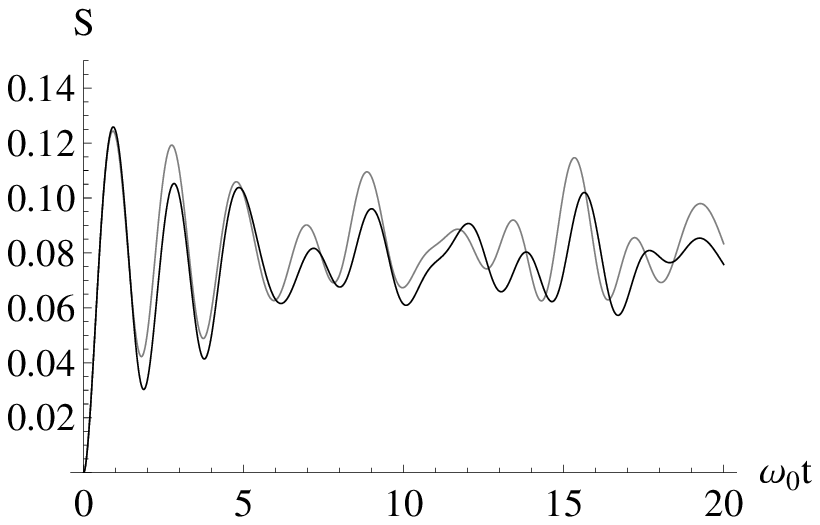}
   {\em \caption{Entropy as a function of time. Solid black: correlator approach to decoherence.
   Solid gray: master equation approach to decoherence. We use $\omega_n/\omega_0 \in [2,3]$, $\lambda/\omega_0^2=1/8$ and $\beta\omega_0=2$.
   \label{fig:SNnonResonance} }}
        \end{center}
   \end{minipage}
\hfill
    \begin{minipage}[t]{.45\textwidth}
        \begin{center}
\includegraphics[width=\textwidth]{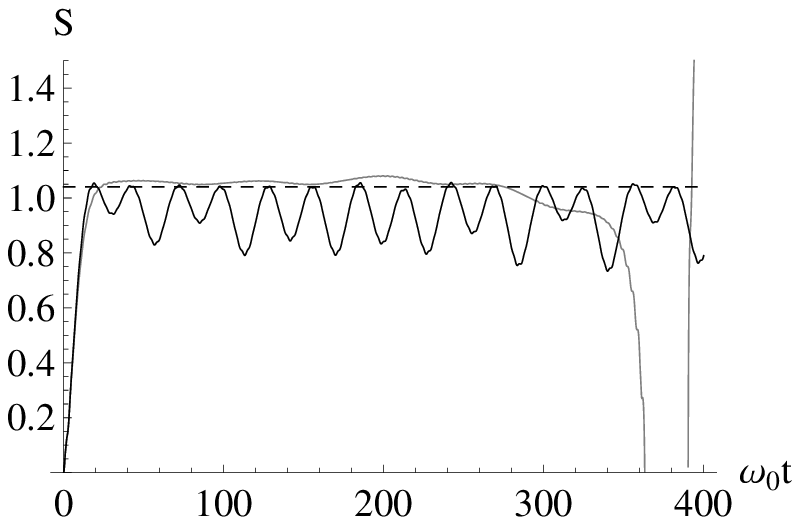}
   {\em \caption{Entropy as a function of time. Solid black: correlator approach to decoherence.
   Solid gray: master equation approach to decoherence. We use \mbox{$\omega_n/\omega_0 \in [9/10,11/10]$,} $\lambda/\omega_0^2=1/40$ and $\beta\omega_0=1$.
   The dashed black line indicates complete thermalisation (perfect decoherence). \label{fig:SNResonance1} }}
        \end{center}
    \end{minipage}
\vskip 0.1cm
    \begin{minipage}[t]{.45\textwidth}
        \begin{center}
\includegraphics[width=\textwidth]{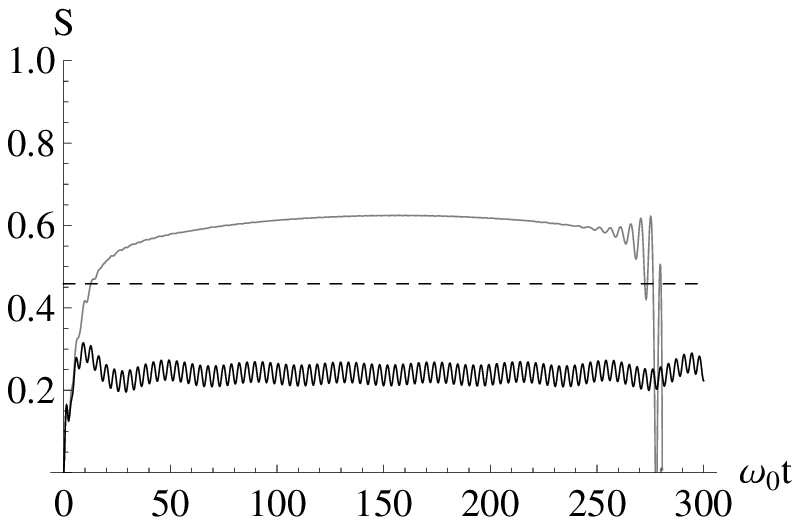}
   {\em \caption{Entropy as a function of time. Solid black: correlator approach to decoherence.
   Solid gray: master equation approach to decoherence. We use \mbox{$\omega_n/\omega_0 = 1+n/50$,} $\lambda/\omega_0^2=3/40$ and $\beta\omega_0=2$.
   \label{fig:SNResonance2} }}
        \end{center}
    \end{minipage}
\hfill
    \begin{minipage}[t]{.45\textwidth}
        \begin{center}
\includegraphics[width=\textwidth]{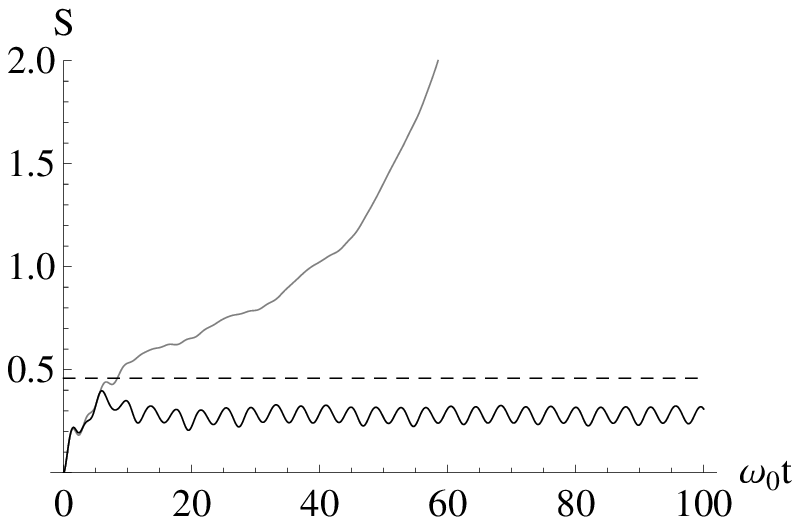}
   {\em \caption{Entropy as a function of time. Solid black: correlator approach to decoherence.
   Solid gray: master equation approach to decoherence. We use \mbox{$\omega_n/\omega_0 = 1+n/100$,} $\lambda/\omega_0^2=3/40$ and $\beta\omega_0=2$.
   \label{fig:SNResonance3} }}
        \end{center}
    \end{minipage}
\vskip 0.1cm
    \begin{minipage}[t]{.45\textwidth}
        \begin{center}
\includegraphics[width=\textwidth]{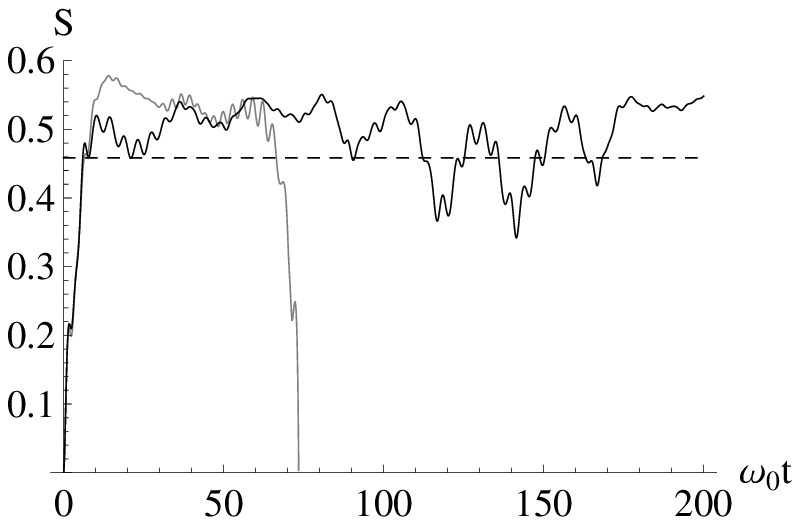}
   {\em \caption{Entropy as a function of time. Solid black: correlator approach to decoherence.
   Solid gray: master equation approach to decoherence. We use \mbox{$\omega_n/\omega_0 \in [3/4,3/2]$,} $\lambda/\omega_0^2=1/16$ and $\beta\omega_0=2$.
   \label{fig:SNResonance4} }}
        \end{center}
    \end{minipage}
\hfill
    \begin{minipage}[t]{.45\textwidth}
        \begin{center}
\includegraphics[width=\textwidth]{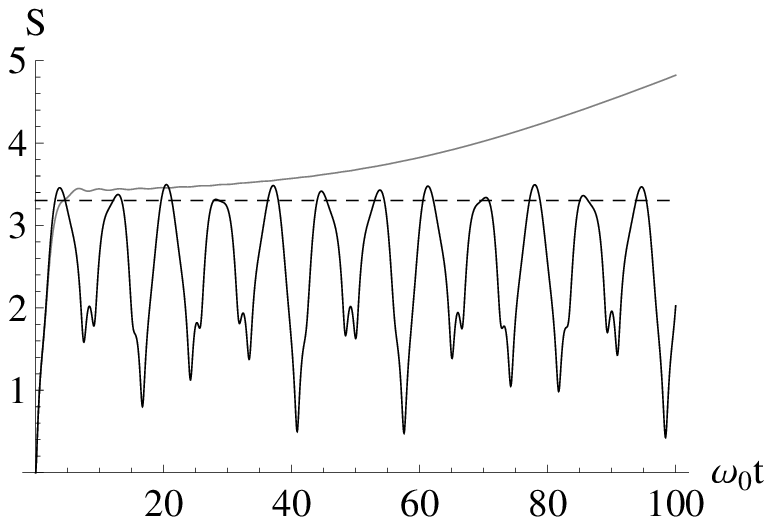}
   {\em \caption{Entropy as a function of time. Solid black: correlator approach to decoherence.
   Solid gray: master equation approach to decoherence. We use \mbox{$\omega_n/\omega_0 \in [19/20,21/20]$,} $\lambda/\omega_0^2=1/10$ and
   $\beta\omega_0=1/10$. At such a high environmental temperature we
   generate a significant amount of entropy.
   \label{fig:SNResonance5} }}
        \end{center}
    \end{minipage}
\end{figure}

We consider $N=50$ environmental oscillators. Let us firstly
present a case where the master equation produces accurate
results. In figure \ref{fig:SNnonResonance} we show the evolution
of the entropy obtained from the master equation (solid gray) and
in our correlator approach to decoherence (solid black). We
randomly generate the environmental frequencies
$\omega_n/\omega_0$, with $1 \leq n \leq N$, in the interval
between 2 and 3. Henceforth, we denote this by $\omega_n/\omega_0
\in [2,3]$. With $\lambda/\omega_0^2=1/8$ this case is clearly
deep in the perturbative regime away from resonance frequencies as
all environmental frequencies are larger than the system
frequency. Only a little entropy is generated so virtually no
decoherence has taken place compared to the thermal value the
entropy can reach in principle if decoherence is effective.
Indeed, given some temperature of the environment, the thermal
value of the entropy provides us with a reasonable estimate for
the upper limit the entropy can reach. For the parameters under
consideration, we have a thermal entropy $S_\mathrm{th} \simeq
0.458$ as $\beta\omega_0=2$, whereas the average generated entropy
is only of the order of $\overline{S}_S \simeq 0.08$. This case is
thus not very different from a $T=0$ evolution. Our system has not
decohered as the environmental frequencies are all larger than the
system frequency, the coupling is weak and the temperature of the
environment is low. In other words, the system and environmental
oscillators are virtually decoupled from each other.

Let us now consider figure \ref{fig:SNResonance1}. Here, we used
$\omega_n/\omega_0 \in [9/10,11/10]$ with $1\leq n \leq N$. These
frequencies are thus randomly spaced around $\omega_0$ in a small
interval. As all frequencies are close to each other, we expect
that thermalisation and decoherence occur swiftly and effectively,
unlike in the previous case we considered. At early times $0 \leq
\omega_0 t \lesssim 15$, we see that the entropy calculated from
the master equation and our correlator approach to decoherence
coincide nicely. This is to be expected as the perturbative master
equation should yield the correct early time evolution. In other
words: it is possible to extract the relevant decoherence time
scale correctly from both the master equation and from our
correlator approach to decoherence.

As entropy or the phase space area quantifies the amount of
decoherence, the rate of change of the phase space area quantifies
the decoherence rate $\Gamma$. We can determine $\Gamma$ at each
moment in time formally by solving:
\begin{equation}\label{DecoherenceRate}
\dot{\Delta}(t) + \Gamma(t) \Delta(t) = 0\,.
\end{equation}
Due to the oscillatory nature of the phase space area as a
function of time, one should average over a conveniently chosen
time interval to sensibly determine for example the decoherence
rate at early times. In this quantum mechanical example it is not
possible to find a simple (time independent) analytic expression
for $\Gamma$, whereas in the field theoretical model we
investigated $\Gamma$ is very well approximated by the tree level
decay rate \cite{Koksma:2009wa}. In \cite{Zurek:2003zz} one can
find a decoherence time for this model that depends on the spatial
coordinates of the density matrix. We do not find an invariant
measure of the decoherence rate of that type.

Now, let us turn our attention to the late time behaviour in
figure \ref{fig:SNResonance1}. Around $\omega_0 t \simeq 280$ the
master equation suddenly destabilises and the entropy becomes ill
defined in the sense that the effective phase space area of the
state becomes negative. One can easily check that the 51
eigenfrequencies one obtains by rotating to the diagonal frame,
analogous to equation (\ref{EOMOscillator4}) for $N=1$, are all
positive so the resulting evolution should be stable. Clearly,
this signals a breakdown of the perturbative master equation.

The latter seems to be a generic feature of the perturbative
master equation in the resonant regime. Let us consider the
figures \ref{fig:SNResonance2}, \ref{fig:SNResonance3},
\ref{fig:SNResonance4} and \ref{fig:SNResonance5}. In figure
\ref{fig:SNResonance2} we use $\omega_n/\omega_0 = 1+n/50$, in
figure \ref{fig:SNResonance3} we use $\omega_n/\omega_0 =
1+n/100$, in figure \ref{fig:SNResonance4}: $\omega_n/\omega_0 \in
[3/4,3/2]$ and in figure \ref{fig:SNResonance5}:
$\omega_n/\omega_0 \in [19/20,21/20]$. Independently on how
precisely one chooses the distribution of environmental
frequencies, the perturbative master equation breaks down. In all
these cases are the eigenfrequencies positive, and we are in all
cases in the perturbative regime, so the evolution of the entropy
should be regular at all times.

The system's entropy should asymptote to the thermal entropy if
thermalisation is complete which we indicate by the dashed black
line in all the figures in this section. A perfectly thermalised
state corresponds to a maximally decohered state and consequently
an imperfectly thermalised state corresponds to a partially
decohered state. Averaging out the oscillations of our unitary
evolution in e.g. figure \ref{fig:SNResonance2} yields an average
value of the entropy that is below this thermal value. This can be
expected given the fact that our coupling is very small
$\lambda/\omega_0^2 =3/40$. Also, one can verify that the energy
is conserved by the evolution as it should.

However, the master equation in figure \ref{fig:SNResonance2}
yields a stationary entropy that is larger than the thermal value,
indicating that the temperature of the system would be higher than
the environmental temperature. Clearly, this does not make sense.
The two most important quantitative measures of decoherence one
hopes to extract out of an experiment or a calculation are the
decoherence rate and the total generated entropy as the latter
tells us how much decoherence has taken place and how classical
the state has become, and the former how fast this state is
reached. We conclude that for this choice of parameters the total
generated entropy at late times does not follow correctly from the
master equation. We checked that this failure of the master
equation is generic. Even in the case of figure
\ref{fig:SNResonance1}, we see that although the asymptote is
roughly equal to the thermal value in the range $25 \lesssim
\omega_0 t \lesssim 280$, it is too high based on our exact
numerical analysis ($\overline{S}_S < S_{\mathrm{th}}$).

The oscillations in the entropy as a function of time that are
present in our exact evolution are not present in the evolution of
the entropy that follows from the master equation. The master
equation thus tends to overdamp oscillations in expectation
values.

Finally, compared to the $N=1$ case we previously considered, we
observe that Poincar\'e's recurrence time has dramatically
increased. For example in figure \ref{fig:SNResonance4} one has to
wait for a very long time before a random fluctuation decreases
the entropy significantly again to values close to its initial
value. Thus, by including more oscillators decoherence becomes
rapidly more irreversible, as one would expect.

\section{Conclusion}
\label{Conclusion}

Decoherence is often studied by considering a perturbative master
equation. This approach suffers from both theoretical and
practical shortcomings. It is unsatisfactory that
$\rho_{\mathrm{red}}$ evolves non-unitarily while the underlying
theory (quantum mechanics or quantum field theory) is unitary. We
are not against non-unitary equations or approximations in
principle, however, one should make sure that the essential
physical features of the system one is describing are kept. The
master equations does not break unitarity correctly, as we have
shown in this paper. On the practical side, the master equation is
so complex that field theoretical questions have barely been
addressed: there does not exist a treatment to take perturbative
interactions properly into account, nor has any reduced density
matrix ever been renormalised. Moreover, as we show in this paper,
the perturbative master equations leads even in simple situations
in quantum mechanics to physically unacceptable behaviour.
Although for early times the entropy increases, for late times the
approach fails and suffers from secular growth as the entropy
continues to grow without bound. This behaviour is generic when at
least some of the environmental oscillators are in the resonant
regime. Of course one could argue that we should have solved the
non-perturbative master equation in the resonant regime. Although
true in principle, the point of this paper is that it is much more
convenient on theoretical and practical grounds to solve for
correlators than for density matrices.

We advocate a novel approach to decoherence based on neglecting
information in observationally unaccessible correlators. If proper
interactions (non-Gaussianities) are taken into account,
neglecting the information stored in higher order non-Gaussian
correlators gives rise to an increase in the Gaussian von Neumann
entropy. In this paper we consider a quadratic model of $N$
coupled simple harmonic oscillators where no non-Gaussianities are
generated. Due to the non-zero coupling however, an increase in
the Gaussian von Neumann entropy can be observed at the expense of
a negative correlation entropy as the total von Neumann entropy is
constant in unitary theories.

In this paper we study a quadratic model where the trace can be
performed, where the perturbative master equation can be solved
and where there is no need to renormalise as we study decoherence
in a quantum mechanical setting. We can thus circumvent the
practical drawbacks that usually prevent us from solving the
perturbative master equation in a renormalised interacting quantum
field theoretical model. In this simple quadratic quantum
mechanical model, we can thus actually compare the master equation
approach to decoherence and our correlator approach to
decoherence. From the numerical analysis we conclude:
\begin{itemize}
\item Away from the resonant regime, where all $\omega_n$ significantly differ from
$\omega_0$, the evolution of the entropy in the master equation
approach and in our correlator approach to decoherence agree
nicely (up to the expected perturbative corrections). The system
decoheres only weakly in this regime however. The two approaches
also agree nicely at very early times such that both approaches
can be used to accurately calculate decoherence rates;
\item In the resonant regime, where a certain number of $\omega_n \simeq
\omega_0$, the perturbative master equation breaks down despite of
the fact that we are still in the perturbative regime and despite
of the fact that all eigenfrequencies are positive. The value to
which the entropy asymptotes in the master equation approach
before the breakdown is reached is generically too high. This
leads to an inaccurate prediction of the amount of decoherence
that has taken place. In some cases, the breakdown occurs already
before this asymptotic state has been reached;
\item The evolution of the entropy in both the resonant and in the non-resonant regime
in our correlator approach to decoherence behaves perfectly
finite. The late time asymptote that is eventually reached either
indicates perfect thermalisation (maximal decoherence) or, for
smaller values of the coupling and a finite number of
environmental oscillators, imperfect thermalisation (imperfect
decoherence).
\end{itemize}
Our correlator approach to decoherence thus provides us with a new
insight in the conventional approach to decoherence. If the
entropy from the master equation agrees with our exact Gaussian
von Neumann entropy up to perturbative corrections, then one
neglects the information stored in the system-environment
correlators in the conventional approach to decoherence too.

There are generally speaking two quantitative results that can be
obtained after performing an experiment or a calculation
concerning decoherence:
\begin{itemize}
\item Decoherence rate;
\item Total amount of decoherence.
\end{itemize}
We argue that one should use the entropy (or, equivalently, the
phase space area) to quantify these two aspects of decoherence.
The decoherence rate $\Gamma$ is the rate at which the phase space
area changes and can formally be obtained from solving equation
(\ref{DecoherenceRate}). The total amount of entropy generated
measures the total amount of decoherence that has occurred. At
late times we have seen that, upon averaging out the oscillations,
a thermal state is maximally decohered. This agrees perfectly with
the point of view where a reduced density matrix whose off
diagonal elements (in the observer's basis) have disappeared
corresponds to a classical state. A (free) thermal density matrix
follows from equations (\ref{density operator: particle2}) and
(\ref{aI:aR:c}) as:
\begin{equation}\label{thermalstate}
\rho_{\mathrm{th}}(x,y;t) = {\cal N}_{\mathrm{th}}(t) \exp\left[ -
\frac{\Delta_{\mathrm{th}}^2}{8\langle
\hat{x}^2\rangle_{\mathrm{th}}}\left(x-y\right)^2 -
\frac{1}{8\langle \hat{x}^2\rangle_{\mathrm{th}}}
\left(x+y\right)^2\right] \,.
\end{equation}
The thermal correlators can be read off for example from equation
(\ref{purethermal_initialconditions}). Clearly, at high
temperatures we have that $\Delta_{\mathrm{th}} \gg 1$ such that
the off diagonal terms in the density matrix have almost
disappeared. The classical limit of quantum mechanics in
\emph{not} classical mechanics but rather classical stochastic
mechanics. Likewise the classical limit of quantum field theory
(in a thermal environment) is not classical field theory but
classical stochastic field theory. This implies that a particular
measurement does not yield a certain predetermined outcome but
rather a certain outcome that is randomly drawn from a probability
distribution function of uncorrelated or not entangled
possibilities\footnote{The Wigner transform of a density matrix is
essentially a Fourier transform with respect to the relative
coordinate $x-y$ in e.g. equation (\ref{thermalstate}) such that
the cross section of the Wigner transform of the thermal density
matrix is a circle. Intuitively, Wigner space provides us (almost)
with a classical stochastic probability distribution function on
phase space.}.

We can learn several things about decoherence in field theories
from our quantum mechanical analysis. Firstly, the reason why the
perturbative master equation fails in the resonant regime is that
the quadratic \textit{coupling} between system and environment is
treated on equal footing as a proper non-Gaussian
\textit{interaction}. In the perturbative master equation, one
attempts to solve the following ``interaction'' in the
``Kadanoff-Baym'' equations:
\begin{equation}\label{KadBaymApprox}
\includegraphics[width=3.0cm]{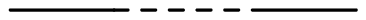} \quad +
\quad
\includegraphics[width=3.0cm]{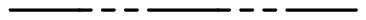} \quad +
\ldots\,.
\end{equation}
Of course, this is not a proper interaction and the 1 particle
irreducible (1PI) self-mass that enters the Kadanoff-Baym
equations\footnote{The Kadanoff-Baym equations are the equations
of motion for the various propagators in the in-in formalism in
quantum field theory that stem from a 2PI effective action such
that the contributing diagrams are 1PI.} due to such a quadratic
coupling vanishes. In this simple quantum mechanical example, one
should therefore just solve for the correlators following from the
von Neumann equation exactly as we showed in this paper. We thus
also accounted for the backreaction from the system on the
environment. The set of Kadanoff-Baym equations is the
sophisticated machinery to properly solve for the propagators in
an interacting quantum field theory. Only the diagrams that follow
from a 2PI effective action contribute. As is well known, the
Kadanoff-Baym equations resum Feynman diagrams and prevent secular
terms from developing. This leads to a stable, thermalised late
time behaviour (see \cite{Berges:2000ur, Berges:2004yj,
Calzetta:book}). Secondly, for $N=1$ and for almost commensurate
eigenfrequencies, we have seen that Poincar\'e's recurrence time
is rather small: the system returns rather quickly to its initial
state. For non-commensurate eigenfrequencies Poincar\'e's
recurrence time increases. For $N=50$ we have seen that
Poincar\'e's recurrence time dramatically increases such that we
have to wait for a much longer time before the system returns to a
state arbitrarily close to its initial state. There are two
complementary ways to interpret this observation. One could take
the point of view that our system has experienced an average
entropy increase, e.g. $0< \overline{S}_S \lesssim
S_{\mathrm{th}}$, where $\overline{S}_S$ is the classical time
average of $S_S(t)$. In other words, the system has experienced a
total amount of decoherence equal to $\overline{S}_S$.
Alternatively, one can say that since the Poincar\'e recurrence
time is still finite, although large, no irreversible process of
decoherence has taken place as the system returns arbitrarily
close to its initial state again. In an interacting field theory
several modes couple due to the loop integrals (hence $N
\rightarrow \infty$) and clearly our Poincar\'e recurrence time
becomes infinite. Hence, the entropy increase has become
irreversible (for all practical purposes) and our system has
decohered.

We hope to generalise the analysis presented here to a quantum
mechanical interaction that comes closer to field theory but is
still free from divergences. In particular, we will consider
entropy generation in a quantum mechanical $\lambda x q^2$ model,
which, when analysed by making use of the master equation, is very
similar to the present $\lambda x q$ model after $q$ has been
integrated out \cite{Hu:1993vs}. The purpose of introducing our
novel approach to decoherence is not to study quadratic quantum
mechanical models, but proper interacting field theories, like in
\cite{Koksma:2009wa}.

\appendix

\section{The Statistical Propagator for $N=1$}
\label{The Statistical Propagator for N1}

Let us now return to the lagrangian~(\ref{action:QM4}) from which
it follows that the equation of motion for the two oscillators can
be written in matrix form as:
\begin{equation}\label{EOMOscillator1}
\frac{\mathrm{d}^2}{\mathrm{d}t^2} \vec q + \Omega \vec q = 0\,,
\end{equation}
where:
\begin{equation}\label{EOMOscillator2}
\Omega = \left(\begin{array}{cc}
                    \omega_0^2 & \lambda_1 \\
                    \lambda_1 & \omega_1^2
                \end{array}
          \right)
\end{equation}
and where we employed the notation $\vec q^{\,\,\mathrm{T}} = ( x
, q)$. We can now easily diagonalise the equation of motion:
\begin{equation}\label{EOMOscillator3}
\frac{\mathrm{d}^2}{\mathrm{d}t^2}\vec{q}_\mathrm{d} +
\Omega_\mathrm{d} \vec{q}_\mathrm{d} = 0 \,,
\end{equation}
where we diagonalised the system by making use of the rotation
matrix $R$:
\begin{equation}\label{EOMOscillator4}
\Omega_\mathrm{d} = R\Omega R^T = \left(\begin{array}{cc}
                    \bar\omega_0^2 & 0 \cr
                    0 & \bar\omega_1^2 \cr
                \end{array}
          \right)
       =  \left(\begin{array}{cc}
                    \frac12(\omega_0^2+\omega_1^2)
     -\frac12\sqrt{(\omega_1^2-\omega_0^2)^2+4\lambda_1^2} & 0\cr
                    0 & \frac12(\omega_0^2+\omega_1^2)
       + \frac12 \sqrt{(\omega_1^2-\omega_0^2)^2+4\lambda_1^2}\cr
                \end{array}
          \right) \,,
\end{equation}
with:
\begin{equation}\label{EOMOscillator5}
R=\left(\begin{array}{cc}
                    \cos(\theta)  & - \sin(\theta)  \cr
                    \sin(\theta)  & \cos(\theta)  \cr
                \end{array}
          \right)\,.
\end{equation}
Moreover, we defined:
\begin{equation}\label{EOMOscillator6}
\vec{q}_\mathrm{d}(t) = \left(\begin{array}{c}
                    \bar x(t)\cr
                    \bar q (t)\cr
                \end{array}
          \right)
        = \left(\begin{array}{c}
                    \cos(\theta)x(t)-\sin(\theta)q(t)\cr
                    \sin(\theta)x(t)+\cos(\theta)q(t) \cr
                \end{array}
          \right)
\,.
\end{equation}
Finally, one can derive that:
\begin{subequations}
\label{EOMOscillator7}
\begin{eqnarray}
\cos(2\theta) &=& \frac{\omega_1^2-\omega_0^2}
{\sqrt{(\omega_1^2-\omega_0^2)^2+4\lambda_1^2}}
\label{EOMOscillator7b}\\
\sin(2\theta) &=& \frac{2\lambda_1}
{\sqrt{(\omega_1^2-\omega_0^2)^2+4\lambda_1^2}}
\label{EOMOscillator7c} \,.
\end{eqnarray}
\end{subequations}
In order for the system to be stable we should have
$\bar{\omega}_0^2\geq 0$, which implies $\lambda_1\leq
\omega_0\omega_1$. When $\lambda_1>\omega_0\omega_1$ the upper
eigenmode corresponds to an inverted harmonic oscillator, $\bar
\omega_{0}^2<0$, and we shall not consider this case here. We can
immediately solve the equation of motion in the diagonal frame:
\begin{equation}
\vec{q}_\mathrm{d} (t) = \left(\begin{array}{c}
                    \bar x(t)\cr
                    \bar q (t)\cr
                \end{array}
          \right)
        = \left(\begin{array}{c}
                    A_0 \cos(\bar{\omega}_0 t) + B_0 \sin(\bar{\omega}_0 t)\cr
                    A_1 \cos(\bar{\omega}_1 t) + B_1 \sin(\bar{\omega}_1 t)\cr
                \end{array}
          \right)
\,, \label{eom:HOs:3}
\end{equation}
We do not yet impose initial conditions for the coefficients in
this solution, but rather first solve for the statistical
propagator for the system:
\begin{eqnarray}
F_x(t;t') &=& \frac{1}{2} \mathrm{Tr} \left[ \hat{\rho}(t_0) \{
\hat{x}(t'),\hat{x}(t) \} \right] = \frac{1}{2} \langle \{
\hat{x}(t'),\hat{x}(t) \} \rangle
\label{statisticalpropagatorsolution1} \\
&=& \frac{1}{2} \Big [ \cos^2(\theta)\langle \{
\hat{\bar{x}}(t'),\hat{\bar{x}}(t) \} \rangle +
\sin^2(\theta)\langle \{ \hat{\bar{q}}(t'),\hat{\bar{q}}(t) \}
\rangle + \cos(\theta)\sin(\theta)\langle \{
\hat{\bar{x}}(t'),\hat{\bar{q}}(t) \} + \{
\hat{\bar{q}}(t'),\hat{\bar{x}}(t) \}\rangle \Big]\nonumber\,.
\end{eqnarray}
Inserting the solution (\ref{eom:HOs:3}) yields the general form
for the statistical propagator:
\begin{eqnarray}
&& \hspace{-0.5cm} F_x(t;t') = \label{statisticalpropagatorsolution2} \\
&& \cos^2(\theta) \Big\{ \langle \hat{A}_0^2\rangle
\cos(\bar{\omega}_0 t)\cos(\bar{\omega}_0 t') + \langle
\hat{B}_0^2\rangle \sin(\bar{\omega}_0 t)\sin(\bar{\omega}_0 t') +
\frac{\langle \{ \hat{A}_0,\hat{B}_0 \} \rangle}{2} \left(
\cos(\bar{\omega}_0 t)\sin(\bar{\omega}_0 t')+ \sin(\bar{\omega}_0
t)\cos(\bar{\omega}_0 t')
\right)\Big\} \nonumber \\
&& +\sin^2(\theta) \Big\{ \langle \hat{A}_1^2\rangle
\cos(\bar{\omega}_1 t)\cos(\bar{\omega}_1 t') + \langle
\hat{B}_1^2\rangle \sin(\bar{\omega}_1 t)\sin(\bar{\omega}_1 t') +
\frac{\langle \{ \hat{A}_1,\hat{B}_1 \} \rangle}{2} \left(
\cos(\bar{\omega}_1 t)\sin(\bar{\omega}_1 t')+
\sin(\bar{\omega}_1 t)\cos(\bar{\omega}_1 t') \right) \! \Big\} \nonumber \\
&& +\frac{\sin(2\theta)}{4} \Big\{ \langle \{ \hat{A}_0,\hat{A}_1
\} \rangle \left( \cos(\bar{\omega}_0 t)\cos(\bar{\omega}_1 t')+
\cos(\bar{\omega}_1 t)\cos(\bar{\omega}_0 t') \right) + \langle \{
\hat{B}_0,\hat{B}_1 \} \rangle \left( \sin(\bar{\omega}_0
t)\sin(\bar{\omega}_1 t')+ \sin(\bar{\omega}_1
t)\sin(\bar{\omega}_0 t') \right) \nonumber \\
&& \qquad\quad + \langle \{ \hat{A}_0,\hat{B}_1 \} \rangle\left(
\cos(\bar{\omega}_0 t)\sin(\bar{\omega}_1 t')+ \sin(\bar{\omega}_1
t)\cos(\bar{\omega}_0 t') \right) + \langle \{ \hat{B}_0,\hat{A}_1
\} \rangle\left( \sin(\bar{\omega}_0 t)\cos(\bar{\omega}_1 t')+
\cos(\bar{\omega}_1 t)\sin(\bar{\omega}_0 t') \right)
\Big\}\nonumber.
\end{eqnarray}
Similarly, we can derive the statistical propagator for the
environment:
\begin{eqnarray}
F_q(t;t') &=& \frac{1}{2} \mathrm{Tr} \left[ \hat{\rho}(t_0) \{
\hat{q}(t'),\hat{q}(t) \} \right] = \frac{1}{2} \langle \{
\hat{q}(t'),\hat{q}(t) \} \rangle
\label{statisticalpropagatorsolution4} \\
&=& \frac{1}{2} \Big [ \sin^2(\theta)\langle \{
\hat{\bar{x}}(t'),\hat{\bar{x}}(t) \} \rangle +
\cos^2(\theta)\langle \{ \hat{\bar{q}}(t'),\hat{\bar{q}}(t) \}
\rangle - \cos(\theta)\sin(\theta)\langle \{
\hat{\bar{x}}(t'),\hat{\bar{q}}(t) \} + \{
\hat{\bar{q}}(t'),\hat{\bar{x}}(t) \}\rangle \Big]\nonumber\,,
\end{eqnarray}
such that we find:
\begin{eqnarray}
&& \hspace{-0.5cm} F_q(t;t') = \label{statisticalpropagatorsolution5} \\
&& \sin^2(\theta) \Big\{ \langle \hat{A}_0^2\rangle
\cos(\bar{\omega}_0 t)\cos(\bar{\omega}_0 t') + \langle
\hat{B}_0^2\rangle \sin(\bar{\omega}_0 t)\sin(\bar{\omega}_0 t') +
\frac{\langle \{ \hat{A}_0,\hat{B}_0 \} \rangle}{2} \left(
\cos(\bar{\omega}_0 t)\sin(\bar{\omega}_0 t')+ \sin(\bar{\omega}_0
t)\cos(\bar{\omega}_0 t')
\right)\Big\} \nonumber \\
&& +\cos^2(\theta) \Big\{ \langle \hat{A}_1^2\rangle
\cos(\bar{\omega}_1 t)\cos(\bar{\omega}_1 t') + \langle
\hat{B}_1^2\rangle \sin(\bar{\omega}_1 t)\sin(\bar{\omega}_1 t') +
\frac{\langle \{ \hat{A}_1,\hat{B}_1 \} \rangle}{2} \left(
\cos(\bar{\omega}_1 t)\sin(\bar{\omega}_1 t')+
\sin(\bar{\omega}_1 t)\cos(\bar{\omega}_1 t') \right) \! \Big\} \nonumber \\
&& -\frac{\sin(2\theta)}{4} \Big\{ \langle \{ \hat{A}_0,\hat{A}_1
\} \rangle \left( \cos(\bar{\omega}_0 t)\cos(\bar{\omega}_1 t')+
\cos(\bar{\omega}_1 t)\cos(\bar{\omega}_0 t') \right) + \langle \{
\hat{B}_0,\hat{B}_1 \} \rangle \left( \sin(\bar{\omega}_0
t)\sin(\bar{\omega}_1 t')+ \sin(\bar{\omega}_1
t)\sin(\bar{\omega}_0 t') \right) \nonumber \\
&& \qquad\quad + \langle \{ \hat{A}_0,\hat{B}_1 \} \rangle\left(
\cos(\bar{\omega}_0 t)\sin(\bar{\omega}_1 t')+ \sin(\bar{\omega}_1
t)\cos(\bar{\omega}_0 t') \right) + \langle \{ \hat{B}_0,\hat{A}_1
\} \rangle\left( \sin(\bar{\omega}_0 t)\cos(\bar{\omega}_1 t')+
\cos(\bar{\omega}_1 t)\sin(\bar{\omega}_0 t') \right)
\Big\}\nonumber.
\end{eqnarray}
Finally, we need the statistical propagator for the
system-environment correlations:
\begin{eqnarray}
F_{xq}(t;t') &=& \frac{1}{2} \mathrm{Tr} \left[ \hat{\rho}(t_0) \{
\hat{x}(t'),\hat{q}(t) \} \right] = \frac{1}{2} \langle \{
\hat{x}(t'),\hat{q}(t) \} \rangle
\label{statisticalpropagatorsolution6} \\
&=& \frac{1}{2} \Big [ \frac{\sin(2\theta)}{2} \left(\langle \{
\hat{\bar{q}}(t'),\hat{\bar{q}}(t) \} \rangle - \langle \{
\hat{\bar{x}}(t'),\hat{\bar{x}}(t) \} \rangle \right) +
\cos^2(\theta)\langle \{ \hat{\bar{x}}(t'),\hat{\bar{q}}(t) \}
\rangle- \sin^2(\theta) \langle \{
\hat{\bar{q}}(t'),\hat{\bar{x}}(t) \}\rangle \Big]\nonumber\,,
\end{eqnarray}
which in turn yields:
\begin{eqnarray}
&& \hspace{-0.5cm} F_{xq}(t;t') = \label{statisticalpropagatorsolution7} \\
&& \frac{\sin(2\theta)}{2} \Big\{ \langle \hat{A}_1^2\rangle
\cos(\bar{\omega}_1 t)\cos(\bar{\omega}_1 t') + \langle
\hat{B}_1^2\rangle \sin(\bar{\omega}_1 t)\sin(\bar{\omega}_1 t') +
\frac{\langle \{ \hat{A}_1,\hat{B}_1 \} \rangle}{2} \left(
\cos(\bar{\omega}_1 t)\sin(\bar{\omega}_1 t')+ \sin(\bar{\omega}_1
t)\cos(\bar{\omega}_1 t') \right) \nonumber \\
&& \qquad\qquad - \langle \hat{A}_0^2\rangle \cos(\bar{\omega}_0
t)\cos(\bar{\omega}_0 t') - \langle \hat{B}_0^2\rangle
\sin(\bar{\omega}_0 t)\sin(\bar{\omega}_0 t') - \frac{\langle \{
\hat{A}_0,\hat{B}_0 \} \rangle}{2} \left( \cos(\bar{\omega}_0
t)\sin(\bar{\omega}_0 t')+ \sin(\bar{\omega}_0
t)\cos(\bar{\omega}_0 t')
\right)\!\Big\}\nonumber \\
&& + \frac{\cos^2(\theta)}{2} \Big\{ \langle \{
\hat{A}_0,\hat{A}_1 \} \rangle \cos(\bar{\omega}_1
t)\cos(\bar{\omega}_0 t') + \langle \{ \hat{B}_0,\hat{B}_1 \}
\rangle \sin(\bar{\omega}_1 t) \sin(\bar{\omega}_0 t')  \nonumber \\
&& \qquad\qquad\qquad\qquad + \langle \{ \hat{A}_0,\hat{B}_1 \}
\rangle \sin(\bar{\omega}_1 t)\cos(\bar{\omega}_0 t')+ \langle \{
\hat{B}_0,\hat{A}_1 \} \rangle \cos(\bar{\omega}_1
t)\sin(\bar{\omega}_0 t') \Big\}\nonumber \\
&& - \frac{\sin^2(\theta)}{2} \Big\{ \langle \{
\hat{A}_0,\hat{A}_1 \} \rangle \cos(\bar{\omega}_0
t)\cos(\bar{\omega}_1 t') + \langle \{ \hat{B}_0,\hat{B}_1 \}
\rangle \sin(\bar{\omega}_0
t)\sin(\bar{\omega}_1 t')\nonumber \\
&& \qquad\qquad\qquad\qquad + \langle \{ \hat{A}_0,\hat{B}_1 \}
\rangle \cos(\bar{\omega}_0 t)\sin(\bar{\omega}_1 t')+ \langle \{
\hat{B}_0,\hat{A}_1 \} \rangle \sin(\bar{\omega}_0
t)\cos(\bar{\omega}_1 t') \Big\}\nonumber ,
\end{eqnarray}
and $F_{qx}(t;t')=F_{xq}(t';t)$. Now, we can impose initial
conditions at $t_0$ on all initial correlations. We can derive:
\begin{subequations}
\label{intialconditions1}
\begin{eqnarray}
\langle \hat{x}^2(t_0)\rangle \!&=& \cos^2(\theta) \langle
\hat{A}_0^{2}\rangle + \sin^2(\theta) \langle \hat{A}_1^{2}\rangle
+\frac{1}{2} \sin (2\theta)\langle\{\hat{A}_0,\hat{A}_1\}\rangle
\label{intialconditions1a}\\
\langle \hat{q}^2(t_0)\rangle \!&=& \sin^2(\theta) \langle
\hat{A}_0^{2}\rangle + \cos^2(\theta) \langle \hat{A}_1^{2}\rangle
-\frac{1}{2} \sin (2\theta)\langle\{\hat{A}_0,\hat{A}_1\}\rangle
\label{intialconditions1b}\\
\langle \hat{x}(t_0)\hat{q}(t_0) \rangle \!&=& \frac{1}{2} \left[
\sin(2\theta)\left( \langle \hat{A}_1^{2}\rangle - \langle
\hat{A}_0^{2}\rangle \right) + \cos(2\theta)
\langle\{\hat{A}_0,\hat{A}_1\}\rangle \right]
\label{intialconditions1c}\\
\langle \hat{p}_x^2(t_0)\rangle \!&=& \cos^2(\theta)
\bar{\omega}_0^2 \langle \hat{B}_0^{2}\rangle +
\sin^2(\theta)\bar{\omega}_1^2 \langle \hat{B}_1^{2}\rangle
+\frac{1}{2} \sin (2\theta) \bar{\omega}_0 \bar{\omega}_1
\langle\{\hat{B}_0,\hat{B}_1\}\rangle
\label{intialconditions1d}\\
\langle \hat{p}_q^2(t_0)\rangle\! &=& \sin^2(\theta)
\bar{\omega}_0^2 \langle \hat{B}_0^{2}\rangle + \cos^2(\theta)
\bar{\omega}_1^2 \langle \hat{B}_1^{2}\rangle -\frac{1}{2} \sin
(2\theta)\bar{\omega}_0 \bar{\omega}_1
\langle\{\hat{B}_0,\hat{B}_1\}\rangle
\label{intialconditions1e}\\
\langle \hat{p}_x(t_0)\hat{p}_q(t_0) \rangle \!&=& \frac{1}{2}
\left[ \sin(2\theta)\left( \bar{\omega}_1^2 \langle
\hat{B}_1^{2}\rangle - \bar{\omega}_0^2 \langle
\hat{B}_0^{2}\rangle \right) + \cos(2\theta) \bar{\omega}_0
\bar{\omega}_1 \langle\{\hat{B}_0,\hat{B}_1\}\rangle \right]
\label{intialconditions1f}\\
\langle \{\hat{x}(t_0),\hat{p}_x(t_0)\} \rangle \!&=&
\cos^2(\theta) \bar{\omega}_0
\langle\{\hat{A}_0,\hat{B}_0\}\rangle + \sin^2(\theta)
\bar{\omega}_1 \langle\{\hat{A}_1,\hat{B}_1\}\rangle +
\frac{1}{2}\sin(2\theta) \left( \bar{\omega}_1
\langle\{\hat{A}_0,\hat{B}_1\}\rangle + \bar{\omega}_0
\langle\{\hat{B}_0,\hat{A}_1\}\rangle \right)
\label{intialconditions1g}\\
\langle \{\hat{q}(t_0),\hat{p}_q(t_0)\} \rangle \!&=&
\sin^2(\theta) \bar{\omega}_0
\langle\{\hat{A}_0,\hat{B}_0\}\rangle + \cos^2(\theta)
\bar{\omega}_1 \langle\{\hat{A}_1,\hat{B}_1\}\rangle-
\frac{1}{2}\sin(2\theta) \left( \bar{\omega}_1
\langle\{\hat{A}_0,\hat{B}_1\}\rangle + \bar{\omega}_0
\langle\{\hat{B}_0,\hat{A}_1\}\rangle \right)
\label{intialconditions1h}\\
\langle \hat{x}(t_0)\hat{p}_q(t_0) \rangle\! &=& \frac{1}{2}
\left[ \frac{\sin(2\theta)}{2} \left( \bar{\omega}_1
\langle\{\hat{A}_1,\hat{B}_1\}\rangle -\bar{\omega}_0
\langle\{\hat{A}_0,\hat{B}_0\}\rangle \right)+ \cos^2(\theta)
\bar{\omega}_1 \langle\{\hat{A}_0,\hat{B}_1\}\rangle
-\sin^2(\theta) \bar{\omega}_0
\langle\{\hat{B}_0,\hat{A}_1\}\rangle \right]
\label{intialconditions1i}\\
\langle \hat{q}(t_0)\hat{p}_x(t_0) \rangle \!&=& \frac{1}{2}
\left[ \frac{\sin(2\theta)}{2} \left( \bar{\omega}_1
\langle\{\hat{A}_1,\hat{B}_1\}\rangle - \bar{\omega}_0
\langle\{\hat{A}_0,\hat{B}_0\}\rangle \right)+ \cos^2(\theta)
\bar{\omega}_0 \langle\{\hat{B}_0,\hat{A}_1\}\rangle
-\sin^2(\theta) \bar{\omega}_1
\langle\{\hat{A}_0,\hat{B}_1\}\rangle \right]
\label{intialconditions1j} \,.
\end{eqnarray}
\end{subequations}
These equations can be inverted to give:
\begin{subequations}
\label{statisticalpropagatorsolution3}
\begin{eqnarray}
\langle \hat{A}_0^2\rangle\! &=& \frac{1}{2} \Big[ \langle
\hat{x}^2(t_0)\rangle +\langle \hat{q}^2(t_0)\rangle
+\cos(2\theta) \left( \langle \hat{x}^2(t_0)\rangle - \langle
\hat{q}^2(t_0)\rangle \right) -2 \sin(2\theta) \langle
\hat{x}(t_0)\hat{q}(t_0)\rangle \Big]
\label{statisticalpropagatorsolution3a}\\
\langle \hat{A}_1^2\rangle\! &=& \frac{1}{2} \Big[ \langle
\hat{x}^2(t_0)\rangle +\langle \hat{q}^2(t_0)\rangle
-\cos(2\theta) \left( \langle \hat{x}^2(t_0)\rangle - \langle
\hat{q}^2(t_0)\rangle \right) +2 \sin(2\theta) \langle
\hat{x}(t_0)\hat{q}(t_0)\rangle \Big]
\label{statisticalpropagatorsolution3b}\\
\langle \{ \hat{A}_0,\hat{A}_1 \}\rangle\! &=& \left( \langle
\hat{x}^2(t_0)\rangle - \langle \hat{q}^2(t_0)\rangle \right)
\sin(2\theta) +2 \cos(2\theta) \langle
\hat{x}(t_0)\hat{q}(t_0)\rangle
\label{statisticalpropagatorsolution3c}\\
\langle \hat{B}_0^2\rangle\! &=& \frac{1}{2 \bar{\omega}_0^2}
\Big[ \langle \hat{p}_x^2(t_0)\rangle +\langle
\hat{p}_q^2(t_0)\rangle +\cos(2\theta) \left( \langle
\hat{p}_x^2(t_0)\rangle - \langle \hat{p}_q^2(t_0)\rangle \right)
-2 \sin(2\theta) \langle \hat{p}_x(t_0)\hat{p}_q(t_0)\rangle \Big]
\label{statisticalpropagatorsolution3d}\\
\langle \hat{B}_1^2\rangle\! &=& \frac{1}{2\bar{\omega}_1^2} \Big[
\langle \hat{p}_x^2(t_0)\rangle +\langle \hat{p}_q^2(t_0)\rangle
-\cos(2\theta) \left( \langle \hat{p}_x^2(t_0)\rangle - \langle
\hat{p}_q^2(t_0)\rangle \right) +2 \sin(2\theta) \langle
\hat{p}_x(t_0)\hat{p}_q(t_0)\rangle \Big]
\label{statisticalpropagatorsolution3e}\\
\langle \{ \hat{B}_0,\hat{B}_1 \}\rangle \!&=&
\frac{1}{\bar{\omega}_0\bar{\omega}_1} \Big[\left( \langle
\hat{p}_x^2(t_0)\rangle - \langle \hat{p}_q^2(t_0)\rangle \right)
\sin(2\theta) +2 \cos(2\theta) \langle
\hat{p}_x(t_0)\hat{p}_q(t_0)\rangle \Big]
\label{statisticalpropagatorsolution3f}\\
\langle \{ \hat{A}_0,\hat{B}_0 \}\rangle \!&=&
\frac{1}{2\bar{\omega}_0} \Big[ \langle \{
\hat{x}(t_0),\hat{p}_x(t_0) \}\rangle + \langle \{
\hat{q}(t_0),\hat{p}_q(t_0) \}\rangle + \cos(2\theta) \left(
\langle \{ \hat{x}(t_0),\hat{p}_x(t_0) \}\rangle - \langle \{
\hat{q}(t_0),\hat{p}_q(t_0) \}\rangle \right) \nonumber \\
&& \qquad - 2 \sin(2\theta) \left( \langle
\hat{q}(t_0)\hat{p}_x(t_0) \rangle + \langle
\hat{x}(t_0)\hat{p}_q(t_0)\rangle \right)\Big]
\label{statisticalpropagatorsolution3g}\\
\langle \{ \hat{A}_1,\hat{B}_1 \} \rangle\! &=&
\frac{1}{2\bar{\omega}_1} \Big[ \langle \{
\hat{x}(t_0),\hat{p}_x(t_0) \}\rangle + \langle \{
\hat{q}(t_0),\hat{p}_q(t_0) \}\rangle - \cos(2\theta) \left(
\langle \{ \hat{x}(t_0),\hat{p}_x(t_0) \}\rangle - \langle \{
\hat{q}(t_0),\hat{p}_q(t_0) \}\rangle \right) \nonumber\\
&& \qquad + 2 \sin(2\theta) \left( \langle
\hat{q}(t_0)\hat{p}_x(t_0) \rangle + \langle
\hat{x}(t_0)\hat{p}_q(t_0) \rangle \right)\Big]
\label{statisticalpropagatorsolution3h}\\
\langle \{ \hat{A}_0,\hat{B}_1 \}\rangle \!&=& \frac{1}{2
\bar{\omega}_1} \Big[ \sin(2\theta) \left( \langle \{
\hat{x}(t_0),\hat{p}_x(t_0) \}\rangle - \langle \{
\hat{q}(t_0),\hat{p}_q(t_0) \}\rangle \right) + 2 \left( \langle
 \hat{x}(t_0)\hat{p}_q(t_0)\rangle - \langle
\hat{q}(t_0)\hat{p}_x(t_0)\rangle   \right)\nonumber \\
&& \qquad + 2 \cos(2\theta) \left( \langle
\hat{x}(t_0)\hat{p}_q(t_0)\rangle + \langle
\hat{q}(t_0)\hat{p}_x(t_0) \rangle \right)\Big]
\label{statisticalpropagatorsolution3i}\\
\langle \{ \hat{B}_0,\hat{A}_1 \}\rangle \!&=& \frac{1}{2
\bar{\omega}_0} \Big[ \sin(2\theta) \left( \langle \{
\hat{x}(t_0),\hat{p}_x(t_0) \}\rangle - \langle \{
\hat{q}(t_0),\hat{p}_q(t_0) \}\rangle \right) -2 \left( \langle
\hat{x}(t_0)\hat{p}_q(t_0) \rangle - \langle
\hat{q}(t_0)\hat{p}_x(t_0) \rangle   \right) \nonumber\\
&& \qquad + 2 \cos(2\theta) \left( \langle
\hat{x}(t_0)\hat{p}_q(t_0)\rangle + \langle
\hat{q}(t_0)\hat{p}_x(t_0) \rangle \right)\Big]
\label{statisticalpropagatorsolution3j}\,.
\end{eqnarray}
\end{subequations}
Note that for example $\hat{x}$ and $\hat{q}$ commute. It might
seem that we brought out the big guns to solve such a simple
problem. This is necessary, however, to generalise the standard
setup to include non-separable initial states. Both Paz and Zurek
\cite{Paz:2000le} and Caldeira and Leggett \cite{Caldeira:1982iu}
for example assume that the density matrix is separable
separability, i.e.:
\begin{equation}\label{separability}
\hat{\rho}(0) = \hat{\rho}_S(0) \otimes  \hat{\rho}_E(0) \,.
\end{equation}
As is apparent from equations (\ref{intialconditions1}) above, we
are now in the position to easily relax this assumption and
generalise to non-separable initial states. Grabert \textit{et
al.} \cite{Grabert:1988yt} and Romero and Paz \cite{Romero:1996bm}
do not assume separable initial conditions in their discussion of
the master equation and consider more general initial states too.

For the ``pure-thermal'' initial conditions given in equation
(\ref{purethermal_initialconditions}), we obtain by means of
equation (\ref{statisticalpropagatorsolution3}):
\begin{subequations}
\label{purethermal_initialconditions2}
\begin{eqnarray}
\langle \hat{A}_0^2\rangle &=& \frac{1}{2} \Bigg[
\frac{1}{2\omega_0} + \frac{1}{2\omega_1}\coth\left(\frac{\beta
\omega_1}{2}\right) +\cos(2\theta) \left( \frac{1}{2\omega_0} -
\frac{1}{2\omega_1}\coth\left(\frac{\beta \omega_1}{2}\right)
\right)\Bigg]
\label{purethermal_initialconditions2a}\\
\langle \hat{A}_1^2\rangle &=& \frac{1}{2} \Bigg[
\frac{1}{2\omega_0} + \frac{1}{2\omega_1}\coth\left(\frac{\beta
\omega_1}{2}\right) -\cos(2\theta) \left( \frac{1}{2\omega_0} -
\frac{1}{2\omega_1}\coth\left(\frac{\beta \omega_1}{2}\right)
\right)\Bigg]
\label{purethermal_initialconditions2b}\\
\langle \{ \hat{A}_0,\hat{A}_1 \}\rangle &=& \sin(2\theta) \left(
\frac{1}{2\omega_0} - \frac{1}{2\omega_1}\coth\left(\frac{\beta
\omega_1}{2}\right) \right)
\label{purethermal_initialconditions2c}\\
\langle \hat{B}_0^2\rangle &=& \frac{1}{2 \bar{\omega}_0^2} \Bigg[
\frac{\omega_0}{2} + \frac{\omega_1}{2}\coth\left(\frac{\beta
\omega_1}{2}\right) +\cos(2\theta) \left( \frac{\omega_0}{2} -
\frac{\omega_1}{2}\coth\left(\frac{\beta \omega_1}{2}\right)
\right)\Bigg]
\label{purethermal_initialconditions2d}\\
\langle \hat{B}_1^2\rangle &=& \frac{1}{2\bar{\omega}_1^2} \Bigg[
\frac{\omega_0}{2} + \frac{\omega_1}{2}\coth\left(\frac{\beta
\omega_1}{2}\right) -\cos(2\theta) \left( \frac{\omega_0}{2} -
\frac{\omega_1}{2}\coth\left(\frac{\beta \omega_1}{2}\right)
\right)\Bigg]
\label{purethermal_initialconditions2e}\\
\langle \{ \hat{B}_0,\hat{B}_1 \}\rangle &=&
\frac{1}{\bar{\omega}_0\bar{\omega}_1} \sin(2\theta) \left(
\frac{\omega_0}{2} - \frac{\omega_1}{2}\coth\left(\frac{\beta
\omega_1}{2}\right) \right)
\label{purethermal_initialconditions2f}\,,
\end{eqnarray}
\end{subequations}
with all other correlators vanishing.

Let us now discuss the time translation invariant states as
initial conditions as discussed in subsection \ref{Initial
Conditions II: Time Translation Invariant States}. In order to
investigate the dependence on the initial conditions, it turns out
to be advantageous to rewrite equation
(\ref{statisticalpropagatorsolution2}) in terms of the average
time and the time difference:
\begin{subequations}
\label{timecoordinates}
\begin{eqnarray}
\tau &=& \frac{1}{2}(t+t') \label{timecoordinatesa}\\
\Delta t &=& t-t' \label{timecoordinatesb} \,.
\end{eqnarray}
\end{subequations}
Making use of several trigonometric identities, this yields:
\begin{eqnarray}\label{statisticalpropagatortimecoordinates}
&& \hspace{-0.6cm} F_x(\tau;\Delta t) = \nonumber \\
&& \phantom{+} \frac{\cos^2(\theta)}{2}\left[ \cos( \bar{\omega}_0
\Delta t)\left[\langle \hat{A}_0^2 \rangle + \langle \hat{B}_0^2
\rangle \right] + \cos( 2 \bar{\omega}_0 \tau)\left[\langle
\hat{A}_0^2 \rangle - \langle \hat{B}_0^2 \rangle \right] + \sin(2
\bar{\omega}_0 \tau)
\langle\{\hat{A}_0,\hat{B}_0\}\rangle \right] \\
&& + \frac{\sin^2(\theta)}{2}\left[ \cos( \bar{\omega}_1 \Delta
t)\left[\langle \hat{A}_1^2 \rangle + \langle \hat{B}_1^2 \rangle
\right] + \cos( 2 \bar{\omega}_1 \tau)\left[\langle \hat{A}_1^2
\rangle - \langle \hat{B}_1^2 \rangle \right] + \sin(2
\bar{\omega}_1 \tau) \langle\{\hat{A}_1,\hat{B}_1\}\rangle \right]
\nonumber\\
&& + \frac{\sin(2\theta)}{4}\Bigg[ \cos (\Delta \bar{\omega} \tau)
\cos(\bar{\omega} \Delta t) \left(
\langle\{\hat{A}_0,\hat{A}_1\}\rangle +
\langle\{\hat{B}_0,\hat{B}_1\}\rangle \right) + \cos
(2\bar{\omega} \tau) \cos(\Delta\bar{\omega} \Delta t/2) \left(
\langle\{\hat{A}_0,\hat{A}_1\}\rangle -
\langle\{\hat{B}_0,\hat{B}_1\}\rangle \right) \nonumber \\
&& \qquad\qquad +\sin (\Delta \bar{\omega} \tau) \cos(\bar{\omega}
\Delta t) \left( \langle\{\hat{B}_0,\hat{A}_1\}\rangle -
\langle\{\hat{A}_0,\hat{B}_1\}\rangle \right) + \sin
(2\bar{\omega} \tau) \cos(\Delta\bar{\omega} \Delta t/2) \left(
\langle\{\hat{B}_0,\hat{A}_1\}\rangle +
\langle\{\hat{A}_0,\hat{B}_1\}\rangle \right)\! \Bigg]\nonumber ,
\end{eqnarray}
where we defined:
\begin{subequations}
\label{frequencycoordinates}
\begin{eqnarray}
\bar{\omega} &=& \frac{1}{2}(\bar{\omega}_0+\bar{\omega}_1) \label{frequencycoordinatesa}\\
\Delta \bar{\omega} &=&\bar{\omega}_0 - \bar{\omega}_1
\label{frequencycoordinatesb} \,.
\end{eqnarray}
\end{subequations}
We now require that the statistical propagator of our system does
not depend on the average time $\tau$, such that:
\begin{subequations}
\label{notimedependence1}
\begin{eqnarray}
\langle \hat{A}_0^2\rangle &=& \langle \hat{B}_0^2\rangle \label{notimedependence1a}\\
\langle \hat{A}_1^2\rangle &=& \langle
\hat{B}_1^2\rangle\label{notimedependence1b} \,,
\end{eqnarray}
\end{subequations}
and all other correlators should vanish. Equation
(\ref{intialconditions1}) thus tells us:
\begin{subequations}
\label{intialconditions2}
\begin{eqnarray}
\langle \hat{x}^2(t_0)\rangle &=& \cos^2(\theta) \langle
\hat{A}_0^{2}\rangle + \sin^2(\theta) \langle \hat{A}_1^{2}\rangle
\label{intialconditions2a}\\
\langle \hat{q}^2(t_0)\rangle &=& \sin^2(\theta) \langle
\hat{A}_0^{2}\rangle + \cos^2(\theta) \langle \hat{A}_1^{2}\rangle
\label{intialconditions2b}\\
\langle \hat{x}(t_0)\hat{q}(t_0) \rangle &=& \frac{1}{2}
\sin(2\theta)\left( \langle \hat{A}_1^{2}\rangle - \langle
\hat{A}_0^{2}\rangle \right)
\label{intialconditions2c}\\
\langle \hat{p}_x^2(t_0)\rangle &=& \cos^2(\theta)
\bar{\omega}_0^2 \langle \hat{A}_0^{2}\rangle +
\sin^2(\theta)\bar{\omega}_1^2 \langle \hat{A}_1^{2}\rangle
\label{intialconditions2d}\\
\langle \hat{p}_q^2(t_0)\rangle &=& \sin^2(\theta)
\bar{\omega}_0^2 \langle \hat{A}_0^{2}\rangle + \cos^2(\theta)
\bar{\omega}_1^2 \langle \hat{A}_1^{2}\rangle
\label{intialconditions2e}\\
\langle \hat{p}_x(t_0)\hat{p}_q(t_0) \rangle &=& \frac{1}{2}
\sin(2\theta)\left( \bar{\omega}_1^2 \langle \hat{A}_1^{2}\rangle
- \bar{\omega}_0^2  \langle \hat{A}_0^{2}\rangle \right)
\label{intialconditions2f}\\
\langle \{\hat{x}(t_0),\hat{p}_x(t_0)\} \rangle &=& 0
\label{intialconditions2g}\\
\langle \{\hat{q}(t_0),\hat{p}_q(t_0)\} \rangle &=& 0
\label{intialconditions2h}\\
\langle \hat{x}(t_0)\hat{p}_q(t_0) \rangle &=& 0
\label{intialconditions2i}\\
\langle \hat{q}(t_0)\hat{p}_x(t_0) \rangle &=& 0
\label{intialconditions2j} \,.
\end{eqnarray}
\end{subequations}
We have thus found a 2-parameter family of initial conditions such
that the statistical propagator for the system does not depend on
the average time. Consequently, the entropy does not depend on the
average time. Let us require that our system is in a pure state
initially, such that:
\begin{subequations}
\label{timeindependentinitialstate1}
\begin{eqnarray}
\langle \hat{x}^2(t_0)\rangle &=& \frac{1}{2\omega_0}
\label{timeindependentinitialstate1a} \\
\langle \hat{p}_x^2(t_0)\rangle &=& \frac{\omega_0}{2}
\label{timeindependentinitialstate1b}\,,
\end{eqnarray}
\end{subequations}
which is equivalent to:
\begin{equation}\label{timeindependentinitialstate2Cond}
\langle \hat{A}_0^2\rangle = \langle \hat{A}_1^2\rangle =
\frac{1}{2\omega_0} \,.
\end{equation}
The other correlators in (\ref{intialconditions2}) can now
trivially be determined:
\begin{subequations}
\label{timeindependentinitialstate2}
\begin{eqnarray}
\langle \hat{q}^2(t_0)\rangle &=& \frac{1}{2\omega_0}
\label{timeindependentinitialstate2a} \\
\langle \hat{p}_q^2(t_0)\rangle &=& \frac{\omega_1^2}{2\omega_0}
\label{timeindependentinitialstate2b} \\
\langle \hat{x}(t_0)\hat{q}(t_0) \rangle &=&  0
\label{timeindependentinitialstate2c} \\
\langle \hat{p}_x(t_0)\hat{p}_q(t_0) \rangle &=&
\frac{\lambda}{2\omega_0} \label{timeindependentinitialstate2d}\,.
\end{eqnarray}
\end{subequations}
Clearly, the environment is in a thermal state at a temperature
dictated by requiring equation (\ref{thermalcondition}):
\begin{equation}
\coth\left( \frac{\beta \omega_1}{2}\right) =
\frac{\omega_1}{\omega_0} \nonumber\,.
\end{equation}
%

\section{Reducing the Density Matrix}
\label{Reducing the Density
Matrix}

In this section we compute the reduced density matrix for our two
coupled simple harmonic oscillators. Starting point for this
derivation is the full density matrix that contains both the
system and environmental degrees of freedom in the Hamiltonian
(\ref{HamiltonianQM1}):
\begin{eqnarray} \label{densitymatrix_full}
\langle x,q| \hat{\rho}(t)| y,r \rangle = \rho(x,q;y,r;t) &=&
\mathcal{N} \exp \Bigg[ - \left(
    \begin{array}{cc}
    x & q
    \end{array}
\right) \left(
    \begin{array}{cc}
    a & d_{1}/2\\
    d_{1}/2 & a_1
    \end{array}
\right) \left(
 \begin{array}{c}
    x \\
    q
    \end{array}
\right)
 - \left(
    \begin{array}{cc}
    y & r
    \end{array}
\right) \left(
    \begin{array}{cc}
    a^{\ast} & d_{1}^{\ast}/2\\
    d_{1}^{\ast}/2 & a_1^{\ast}
    \end{array}
\right) \left(
 \begin{array}{c}
    y \\
    r
    \end{array}
\right) \nonumber \\
&& \qquad\qquad + 2 \left(
    \begin{array}{cc}
    x & q
    \end{array}
\right) \left(
    \begin{array}{cc}
    c & e_1/2\\
    e_1^{\ast}/2 & c_1
    \end{array}
\right) \left(
 \begin{array}{c}
    y \\
    r
    \end{array}
\right) \Bigg]  \,,
\end{eqnarray}
where $c$ and $c_1$ are real such that there are precisely ten
degrees of freedom in this density matrix. Of course the density
matrix is hermitian: $\rho(x,q;y,r;t)  = \rho^{\ast}(y,r;x,q;t) $.
The normalisation constant $\mathcal{N}$ is determined by
requiring $\mathrm{Tr}[\hat{\rho}(t)]=1$, yielding:
\begin{equation} \label{densitymatrix_full_normalization}
\mathcal{N} = \frac{\sqrt{4(a^{\mathrm{R}} - c)(a_1^{\mathrm{R}} -
c_1) - (d_1^{\mathrm{R}} - e_1^{\mathrm{R}})^2}}{\pi}
\end{equation}
where, for normalisability we required:
\begin{subequations}
\label{densitymatrix_full_normalization2}
\begin{eqnarray}
a_1^{\mathrm{R}} - c_1 &>& 0
\label{densitymatrix_full_normalization2a}\\
4(a^{\mathrm{R}} - c)(a_1^{\mathrm{R}} - c_1)  &>&
(d_1^{\mathrm{R}} - e_1^{\mathrm{R}})^2
\label{densitymatrix_full_normalization2b}\,.
\end{eqnarray}
\end{subequations}
The unitary dynamics of the ten degrees of freedom in this density
matrix is governed by the von Neumann equation
(\ref{vNeumannEquation}). Of course we can trace the density
matrix which yields:
\begin{equation} \label{reduceddensitymatrix2}
\hat{\rho}_{\mathrm{red}}(x;y;t)=
\int_{-\infty}^{\infty}\mathrm{d} q \, \rho(x,q;y,q;t) =
\tilde{\mathcal{N}} \exp\left[ -\tilde{a}
x^2-\tilde{b}y^2+2\tilde{c}xy \right] \,,
\end{equation}
where:
\begin{subequations}
\label{reduceddensitymatrix3}
\begin{eqnarray}
\tilde{a} &=& a - \frac{1}{8(a_1^{\mathrm{R}} - c_1)}( d_1 -
e_1)^2
\label{reduceddensitymatrix3a}\\
\tilde{b} &=& a^{\ast} - \frac{1}{8(a_1^{\mathrm{R}} - c_1)}(
d_1^{\ast} - e_1^{\ast})^2
\label{reduceddensitymatrix3b} \\
\tilde{c} &=& c + \frac{|d_1-e_1|^{2}}{8(a_1^{\mathrm{R}} -
c_1)} \label{reduceddensitymatrix3c}\\
\tilde{\mathcal{N}} &=& \mathcal{N} \sqrt{\frac{\pi}{
2(a_1^{\mathrm{R}} - c_1)}} = \sqrt{\frac{2(a_{\mathrm{R}} -
c)}{\pi} - \frac{ (d_1^{\mathrm{R}} -
e_1^{\mathrm{R}})^2}{2\pi(a_1^{\mathrm{R}} - c_1)}}
\label{reduceddensitymatrix3d} \,.
\end{eqnarray}
\end{subequations}
The total von Neumann entropy can now straightforwardly be
obtained using the replica trick \cite{Koksma:2010zi}:
\begin{equation}
S_{\rm red}(t) =
\frac{\tilde{\Delta}+1}{2}\ln\left(\frac{\tilde{\Delta}+1}{2}\right)
-
\frac{\tilde{\Delta}-1}{2}\ln\left(\frac{\tilde{\Delta}-1}{2}\right)\,,
\label{entropyreduced}
\end{equation}
where:
\begin{equation}
\tilde{\Delta}^2 = \frac{\tilde{a}_{\mathrm{R}} +
\tilde{c}}{\tilde{a}_{\mathrm{R}} - \tilde{c}} \,.
\label{entropyreduced2}
\end{equation}
Although this expression is the final answer for the entropy
generated by tracing out the environmental degrees of freedom, it
is not in a convenient form to study the dynamics. All we have
done is relate the von Neumann entropy to the coefficients in the
full density matrix. The strategy is as follows: the ten degrees
of freedom in the density matrix can straightforwardly obtained by
numerically solving the full density matrix. If we insert the
Ansatz for the density matrix (\ref{densitymatrix_full}) in the
von Neumann equation, we find the following set of differential
equations:
\begin{subequations}
\label{vNeumannequation2}
\begin{eqnarray}
\frac{\mathrm{d}a_{\mathrm{R}}}{\mathrm{d}t} &=& 4 a_{\mathrm{I}}
a_{\mathrm{R}} +  \left(d_1^{\mathrm{R}} d_1^{\mathrm{I}} -
e_1^{\mathrm{R}}
e_1^{\mathrm{I}}\right )\label{vNeumannequation2a}\\
\frac{\mathrm{d}a_{\mathrm{I}}}{\mathrm{d}t} &=&
\frac{\omega_0^2}{2} + 2 \left( a_{\mathrm{I}}^2 -
a_{\mathrm{R}}^2 + c^2 \right) -\frac{1}{2} \left(
(d_1^{\mathrm{R}})^2 - (e_1^{\mathrm{R}})^2 - (d_1^{\mathrm{I}})^2
+ (e_1^{\mathrm{I}})^2 \right) \label{vNeumannequation2b}\\
\frac{\mathrm{d}c}{\mathrm{d}t} &=& 4 a_{\mathrm{I}} c -  \left(
d_1^{\mathrm{R}} e_1^{\mathrm{I}} - e_1^{\mathrm{R}}
d_1^{\mathrm{I}}\right ) \label{vNeumannequation2c}\\
\frac{\mathrm{d}a_1^{\mathrm{R}}}{\mathrm{d}t} &=& 4
a_{1}^{\mathrm{I}} a_1^{\mathrm{R}} +  \left(d_1^{\mathrm{R}}
d_1^{\mathrm{I}} - e_1^{\mathrm{R}} e_1^{\mathrm{I}}\right
)\label{vNeumannequation2d}\\
\frac{\mathrm{d}a_1^{\mathrm{I}}}{\mathrm{d}t} &=&
\frac{\omega_1^2}{2} + 2 \left( (a_1^{\mathrm{I}})^2 -
(a_1^{\mathrm{R}})^2 + c_1^2 \right) -\frac{1}{2} \left(
(d_1^{\mathrm{R}})^2 - (e_1^{\mathrm{R}})^2 - (d_1^{\mathrm{I}})^2
+ (e_1^{\mathrm{I}})^2 \right) \label{vNeumannequation2e}\\
\frac{\mathrm{d}c_1}{\mathrm{d}t} &=& 4 a_1^{\mathrm{I}} c_1 +
\left( d_1^{\mathrm{R}} e_1^{\mathrm{I}} + e_1^{\mathrm{R}}
d_1^{\mathrm{I}}\right ) \label{vNeumannequation2f}\\
\frac{\mathrm{d}d_1^{\mathrm{R}}}{\mathrm{d}t} &=& 2 \left[ \left(
a_{\mathrm{R}} + a_1^{\mathrm{R}}\right)d_1^{\mathrm{I}} + \left(
a_{\mathrm{I}} + a_1^{\mathrm{I}}\right)d_1^{\mathrm{R}}+
\left( c - c_1\right)e_1^{\mathrm{I}}\right] \label{vNeumannequation2g}\\
\frac{\mathrm{d}d_1^{\mathrm{I}}}{\mathrm{d}t} &=& \lambda + 2
\left[ -\left( a_{\mathrm{R}} +
a_1^{\mathrm{R}}\right)d_1^{\mathrm{R}} + \left( a_{\mathrm{I}} +
a_1^{\mathrm{I}}\right)d_1^{\mathrm{I}}+ \left( c +
c_1\right)e_1^{\mathrm{R}} \right] \label{vNeumannequation2h} \\
\frac{\mathrm{d}e_1^{\mathrm{R}}}{\mathrm{d}t} &=& 2 \left[ \left(
a_{\mathrm{R}} - a_1^{\mathrm{R}}\right)e_1^{\mathrm{I}} + \left(
a_{\mathrm{I}} + a_1^{\mathrm{I}}\right)e_1^{\mathrm{R}}+
\left( c + c_1\right)d_1^{\mathrm{I}}\right] \label{vNeumannequation2i} \\
\frac{\mathrm{d}e_1^{\mathrm{I}}}{\mathrm{d}t} &=&  2 \left[
-\left( a_{\mathrm{R}}- a_1^{\mathrm{R}}\right)e_1^{\mathrm{R}} +
\left( a_{\mathrm{I}} + a_1^{\mathrm{I}}\right)e_1^{\mathrm{I}}+
\left( c - c_1\right)d_1^{\mathrm{R}} \right]
\label{vNeumannequation2j} \,.
\end{eqnarray}
\end{subequations}
For pure and thermal initial states as previously discussed (the
system is in a pure state, the environment in a thermal state),
one can straightforwardly show that the reduced Gaussian von
Neumann entropy that results from equation (\ref{entropyreduced})
coincides precisely with the Gaussian von Neumann entropy $S_S(t)$
defined in equation (\ref{vNeumannEntropy6}). We can thus confirm
the identity (\ref{entropyequality}) in an explicit model:
\begin{equation}
S_{S}(t) = S_{\mathrm{vN}}^{\mathrm{red}}(t)\nonumber\,.
\end{equation}

\end{document}